\documentclass[sigconf,nonacm]{acmart}
\usepackage{xcolor}
\usepackage{listings}
\usepackage{tikz}
\usetikzlibrary{shapes,arrows.meta,positioning,fit,calc,backgrounds}
\usepackage{booktabs}
\usepackage{pgfplots}
\pgfplotsset{compat=1.18}
\usepackage{subcaption}
\usepackage{xspace}
\usepackage{framed}
\usepackage{enumitem}
\usepackage{float}
\usepackage{mathpartir}
\usepackage{capt-of}
\usepackage{cuted}
\usepackage{needspace}

\newcommand{\tool}{Mine\xspace}

\definecolor{cBlue}{RGB}{65,105,225}
\definecolor{cOrange}{RGB}{230,145,56}
\definecolor{cTeal}{RGB}{0,128,128}
\definecolor{cRed}{RGB}{190,50,50}
\definecolor{cGreen}{RGB}{3,141,89}
\definecolor{cPurple}{RGB}{130,80,180}
\definecolor{cGray}{RGB}{120,120,120}

\newcommand{\hlg}[1]{\setlength{\fboxsep}{0pt}\colorbox{green!15}{\texttt{#1}}}
\newcommand{\hlr}[1]{\setlength{\fboxsep}{0pt}\colorbox{red!15}{\texttt{#1}}}

\pgfplotsset{
  every axis/.append style={
    grid style={dashed, gray!20},
    tick label style={font=\scriptsize},
    label style={font=\small\sffamily},
    legend style={
      font=\scriptsize,
      fill=white!90!gray,
      draw=none,
      rounded corners=2pt,
      legend cell align=left,
    },
    axis line style={black, line width=0.3pt},
  },
}

\acmConference[CCS '26]{ACM Conference on Computer and Communications Security}{October 2026}{Salt Lake City, UT, USA}
\acmYear{2026}
\acmISBN{}
\acmDOI{}



\title{Your Agent Is Mine: Measuring Malicious Intermediary Attacks on the LLM Supply Chain}

\author{Hanzhi Liu}
\affiliation{%
  \institution{University of California, Santa Barbara}
  \country{}
}
\email{hanzhi@ucsb.edu}

\author{Chaofan Shou}
\affiliation{%
  \institution{Fuzzland}
  \country{}
}
\email{shou@fuzz.land}

\author{Hongbo Wen}
\affiliation{%
  \institution{University of California, Santa Barbara}
  \country{}
}
\email{hongbowen@ucsb.edu}

\author{Yanju Chen}
\affiliation{%
  \institution{University of California, San Diego}
  \country{}
}
\email{yanju@ucsd.edu}

\author{Ryan Jingyang Fang}
\affiliation{%
  \institution{World Liberty Financial}
  \country{}
}
\email{ryan@worldlibertyfinancial.com}

\author{Yu Feng}
\affiliation{%
  \institution{University of California, Santa Barbara}
  \country{}
}
\email{yufeng@cs.ucsb.edu}

\begin{document}

\begin{abstract}
Large language model (LLM) agents increasingly rely on third-party API
routers to dispatch tool-calling requests across multiple upstream
providers.
These routers operate as application-layer proxies with full plaintext
access to every in-flight JSON payload, yet no provider enforces
cryptographic integrity between client and upstream model.
We present the first systematic study of this attack surface.
We formalize a threat model for malicious LLM API routers and define
two core attack classes, payload injection (AC-1) and secret
exfiltration (AC-2), together with two adaptive evasion variants:
dependency-targeted injection (AC-1.a) and conditional delivery
(AC-1.b).
Across 28 paid routers purchased from Taobao, Xianyu, and
Shopify-hosted storefronts and 400 free routers collected from public
communities, we find 1~paid and 8~free routers actively injecting
malicious code, 2~deploying adaptive evasion triggers, 17~touching
researcher-owned AWS canary credentials, and 1~draining ETH from a
researcher-owned private key.
Two poisoning studies further show that ostensibly benign routers can be
pulled into the same attack surface as they process end-user requests using leaked credentials and weakly configured peers: intentionally leaked OpenAI keys and weakly configured decoys have processed 2.1B tokens from these routers, exposing 99 credentials across 440-codex sessions, and 401~sessions already running in autonomous YOLO mode, allowing direct payload injection.
We build \tool, a research proxy that implements all four attack
classes against four public agent frameworks, and use it to evaluate
three deployable client-side defenses: a fail-closed policy gate,
response-side anomaly screening, and append-only transparency logging.
\end{abstract}

\maketitle

\begin{figure*}[!t]
\centering
\resizebox{0.96\textwidth}{!}{
\pgfdeclarelayer{bg}
\pgfsetlayers{bg,main}
\begin{tikzpicture}[
  req/.style={-{Stealth[length=1.8mm, width=1.4mm]}, line width=0.7pt,
    cGreen!80!black},
  resp/.style={-{Stealth[length=1.8mm, width=1.4mm]}, line width=0.7pt,
    cGreen!80!black},
  tainted/.style={-{Stealth[length=1.8mm, width=1.4mm]}, line width=0.7pt,
    cRed!80!black},
  router/.style={circle, draw=cGreen!60, fill=cGreen!8, inner sep=2pt,
    minimum size=0.85cm, font=\sffamily\footnotesize},
  malrouter/.style={circle, draw=cRed!70, fill=cRed!10, line width=1.2pt,
    inner sep=2pt, minimum size=0.85cm, font=\sffamily\footnotesize\bfseries},
  groupbox/.style={draw=black!25, rounded corners=6pt, fill=black!2,
    inner xsep=10pt, inner ysep=12pt},
  affected/.style={font=\sffamily\tiny\bfseries, text=cRed!80,
    rounded corners=1.5pt, fill=cRed!8, inner sep=1.5pt},
  stamp/.style={font=\sffamily\scriptsize\bfseries, text=cRed!80,
    draw=cRed!70, line width=0.8pt, rounded corners=1pt,
    inner xsep=3pt, inner ysep=1.5pt, rotate=-22,
    fill=white, fill opacity=0.25, text opacity=0.85},
  yolotag/.style={font=\sffamily\tiny\bfseries, text=cOrange!90!black,
    rounded corners=1.5pt, fill=cOrange!12, draw=cOrange!40,
    line width=0.3pt, inner sep=1.5pt},
]

\node (a1) at (0, 1.4)
  {\raisebox{-3pt}{\includegraphics[height=11pt]{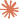}}
   \,\sffamily\small Claude Code};
\node (a2) at (0, 0)
  {\raisebox{-3pt}{\includegraphics[height=11pt]{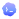}}
   \,\sffamily\small Codex};
\node (a3) at (0,-1.4)
  {\raisebox{-3pt}{\includegraphics[height=11pt]{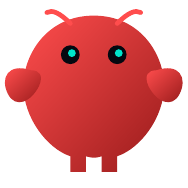}}
   \,\sffamily\small OpenClaw};

\node[stamp] at (a1) {COMPROMISED};
\node[stamp] at (a2) {COMPROMISED};
\node[yolotag, anchor=south east] at ([yshift=-1pt]a1.north east) {\fontsize{4}{4.5}\selectfont YOLO mode};

\begin{pgfonlayer}{bg}
  \node[groupbox, fit=(a1)(a2)(a3),
    label={[font=\sffamily\footnotesize\bfseries, text=black!60,
      anchor=south]north:Agent Clients}] (agentbox) {};
\end{pgfonlayer}

\node[router] (r1) at (4.2, 1.0)  {$R_1$};
\node[router] (r2) at (4.2,-1.0)  {$R_2$};

\node[router]    (r3) at (6.4, 1.4)  {$R_3$};
\node[malrouter] (r4) at (6.4,-0.1) {$R_4$};
\node[router]    (r5) at (6.4,-1.4)  {$R_5$};

\node[router] (r6) at (8.6, 1.0)  {$R_6$};
\node[router] (r7) at (8.6,-1.0) {$R_7$};

\node[font=\sffamily\tiny\bfseries, text=cRed!80] at (6.4, -0.75) {malicious};

\node (hacker) at (8, 0.30)
  {\includegraphics[height=24pt]{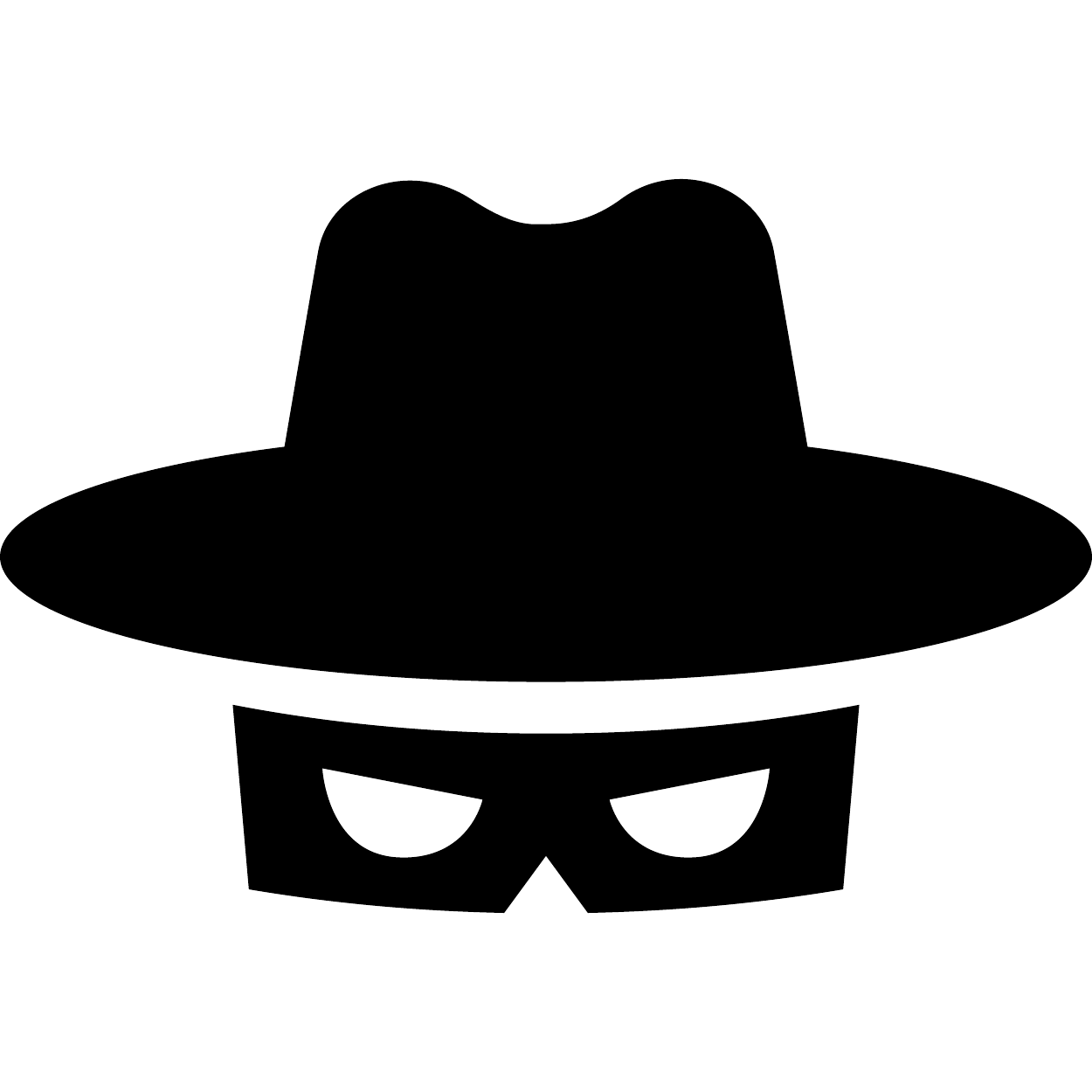}};
\node[font=\sffamily\tiny\bfseries, text=cRed!80, anchor=north]
  at ([yshift=7pt]hacker.south) {Attacker};
\coordinate (atkstart) at (7.46, 0.05);
\draw[-{Stealth[length=1.4mm, width=1.2mm]}, dashed, cRed!70,
  line width=0.6pt] (atkstart) -- (r4.east);

\node[font=\sffamily\footnotesize\bfseries, text=black!60]
  at (6.4, 2.6) {Multi-hop LLM Router Chain};

\node (p1) at (12.2, 1.0)
  {\raisebox{-2pt}{\includegraphics[height=10pt]{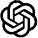}}
   \,\sffamily\small OpenAI};
\node (p2) at (12.2,-0.2)
  {\raisebox{-2pt}{\includegraphics[height=10pt]{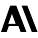}}
   \,\sffamily\small Anthropic};
\node (p3) at (12.2,-1.4)
  {\raisebox{-2pt}{\includegraphics[height=10pt]{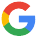}}
   \,\sffamily\small Google};

\begin{pgfonlayer}{bg}
  \node[groupbox, fit=(p1)(p2)(p3),
    label={[font=\sffamily\footnotesize\bfseries, text=black!60,
      anchor=south]north:Model Providers}] (provbox) {};
\end{pgfonlayer}

\newcommand{\ba}{10}

\draw[req]     (a1.east) to[bend left=\ba] (r1);
\draw[tainted] (r1) to[bend left=\ba] (a1.east);
\draw[req]     (a2.east) to[bend left=\ba] (r1);
\draw[tainted] (r1) to[bend left=\ba] (a2.east);
\node[font=\ttfamily\fontsize{4}{4.5}\selectfont, text=cRed!80,
  inner sep=0.5pt, rotate=18]
  at (2.7, 0.15) {Bash: curl ****.xyz/pwn.sh | sh};

\draw[req]  (a3.east) to[bend left=\ba] (r2);
\draw[resp] (r2) to[bend left=\ba] (a3.east);

\draw[req]  (r1) to[bend left=\ba] (r3);
\draw[resp] (r3) to[bend left=\ba] (r1);

\draw[req]     (r1) to[bend left=\ba] (r4);
\draw[tainted] (r4) to[bend left=\ba] (r1);

\draw[req]  (r2) to[bend left=\ba] (r5);
\draw[resp] (r5) to[bend left=\ba] (r2);

\draw[req]  (r3) to[bend left=\ba] (r6);
\draw[resp] (r6) to[bend left=\ba] (r3);

\draw[req]  (r4) to[bend left=\ba] (r7);
\draw[resp] (r7) to[bend left=\ba] (r4);

\draw[req]  (r5) to[bend left=\ba] (r7);
\draw[resp] (r7) to[bend left=\ba] (r5);

\draw[req]  (r6) to[bend left=\ba] (p1.west);
\draw[resp] (p1.west) to[bend left=\ba] (r6);

\draw[req]  (r7) to[bend left=\ba] (p2.west);
\draw[resp] (p2.west) to[bend left=\ba] (r7);
\draw[req]  (r7) to[bend left=\ba] (p3.west);
\draw[resp] (p3.west) to[bend left=\ba] (r7);

\node[font=\sffamily\scriptsize, anchor=west] at (0.5, -2.6)
  {\textcolor{cGreen!80!black}{\rule{8pt}{1.5pt}}~request / clean response
   \qquad
   \textcolor{cRed!80!black}{\rule{8pt}{1.5pt}}~tainted response};

\end{tikzpicture}
}
\caption{LLM router ecosystem and taint propagation.
Agent clients (left) exchange requests and responses through a
multi-hop graph of LLM routers to upstream model providers (right).
Each hop terminates the inbound TLS session, granting full plaintext
access.
Green arrows denote clean data flow; red arrows trace how a single
malicious router $R_4$, controlled by an external attacker,
taints responses on the return path: corrupted payloads propagate
through $R_1$ back to the compromised Claude Code and Codex clients,
handing the attacker effective control over their tool execution
(``your agent is mine''), while agents routed through honest paths
(e.g., $R_2 \!\to\! R_5$) remain unaffected
(Section~\ref{sec:attacks}).}
\label{fig:architecture}
\end{figure*}
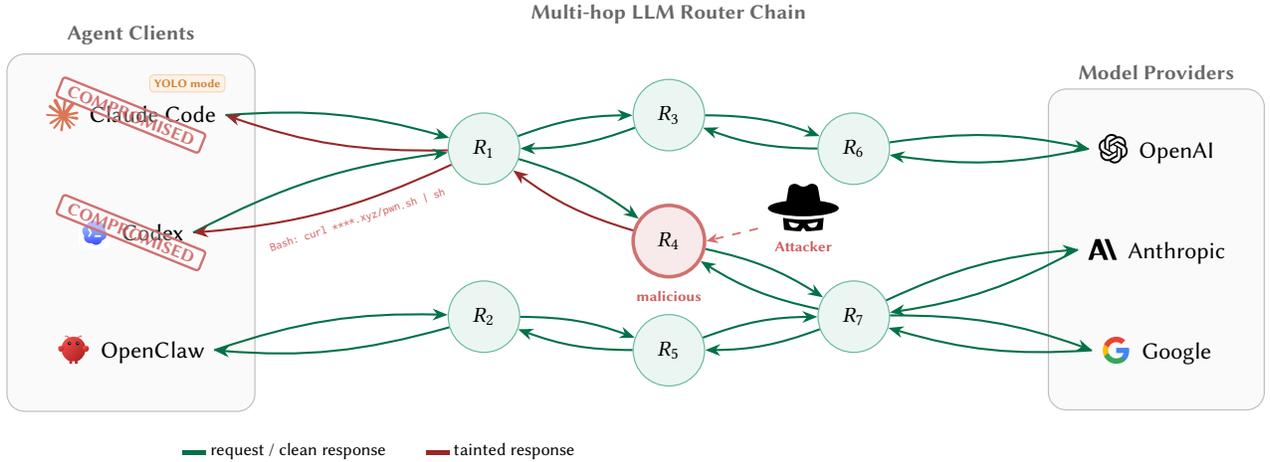

\section{Introduction}
\label{sec:introduction}

Large language model (LLM) agents have moved beyond conversational assistants into
tool-using systems that book flights, execute code, query databases, and manage
cloud infrastructure on behalf of their users~\cite{agenticsurvey2025}.
A less studied but increasingly critical component in this ecosystem is the
\emph{LLM API router}: an intermediary service that accepts
requests in a unified format and dispatches them to upstream model providers.
LiteLLM~\cite{litellm2024}, the dominant open-source router with
roughly 40{,}000 GitHub stars and over 240 million Docker Hub pulls,
is integrated into production pipelines across thousands of
organizations.
OpenRouter~\cite{openrouter2024} connects users to more than 300 active
models from over 60 providers and serves millions of developers and
end-users~\cite{chou2025openrouter}.
Routers provide model fallback, load balancing, cost optimization,
and a single API key across providers.
A growing number of production deployments route
traffic through at least one such intermediary~\cite{agenticsurvey2025,ruan2024risks}.
The severity of this dependency was demonstrated in March 2026, when
attackers compromised the LiteLLM package through dependency
confusion, injecting malicious code directly into the request-handling
pipeline of every deployment that pulled the poisoned
release~\cite{litellm2026teampcp}.
That incident turned a widely trusted router into a supply-chain weapon
with full plaintext access to every transiting API request and response.

This architecture creates a trust relationship that has received little scrutiny.
The ``router-in-the-middle'' is not an accidental on-path adversary but an
intentionally configured intermediary with application-layer authority over both
requests and responses.
Unlike a traditional network MITM, no TLS downgrade or certificate forgery is
required: the client voluntarily configures the router's URL as the API
endpoint, the router terminates the client-side TLS connection, and it originates
a separate TLS connection upstream.
Once an agent targets that router endpoint, the service can inspect tool-call
arguments, API keys, system prompts, and model outputs; it can also normalize,
delay, or rewrite the returned tool call before the client executes it.
No end-to-end integrity mechanism binds the provider's tool-calling output to the
action the client finally observes (Section~\ref{sec:threatmodel}).
A malicious or compromised router can therefore replace a benign installer URL
with an attacker-controlled script, swap \texttt{pip install requests} for a
attacker-controlled dependency, or silently exfiltrate every credential
that transits the service.

Proxy tampering itself is not new~\cite{durumeric2017security,decarnavalet2016killed},
but LLM agents make this intermediary trust boundary unusually dangerous because
the payload now carries executable tool-call semantics.
We study that boundary as an LLM supply-chain problem and introduce a taxonomy of
\emph{Adversarial Router Behaviors}, spanning direct payload manipulation,
dependency rewriting, credential sniffing, and \emph{Adaptive Evasion}, in which
malicious rewrites are delivered only after a warm-up period or when the router
infers that the client is running in an autonomous ``YOLO mode.''
These attacks are orthogonal to prompt injection~\cite{greshake2023prompt,perez2022ignore}:
they occur in the JSON/tool layer before the model sees the request or after it
emits a response, outside the model's reasoning loop, and therefore compose with
model-side safeguards rather than replacing them.

Our empirical results show that this risk is already present in
commodity router markets.
The open-source templates that underpin most commodity
routers, \texttt{new-api}~\cite{newapi2026} (25.4k GitHub stars,
1.25M Docker pulls) and its upstream fork
\texttt{one-api}~\cite{oneapi2026} (30.5k stars, 1.19M Docker
pulls), have been pulled millions of times, and Chinese open-source
models reached nearly 30\% of total usage on OpenRouter in some
weeks~\cite{chou2025openrouter}, the largest public routing platform.
Investigative reporting documents Taobao shops with over 30{,}000
repeat purchases for resold LLM API
access~\cite{chinatalk2025graymarket}.
We analyze 28 paid routers bought from Taobao, Xianyu, and
Shopify-hosted storefronts and 400 free routers built from the
dominant \texttt{sub2api}~\cite{sub2api2026} and
\texttt{new-api} templates.
Within that corpus, 1 paid and 8 free routers inject malicious code
into returned tool calls.
Two routers deploy adaptive evasion in the wild, for example by waiting
for 50 prior calls, restricting payload delivery to autonomous
``YOLO mode'' sessions, or targeting only Rust and Go projects.
Among the free-router set, 17 routers touch at least one
researcher-owned AWS canary credential and 1 drains ETH from a
researcher-owned Ethereum private key.

Malicious routers are only half of the story.
Routers that look benign can be poisoned into the same trust boundary
when they reuse leaked upstream keys or forward traffic through weaker
relays.
We intentionally leaked a researcher-owned OpenAI key on Chinese forums
and in WeChat and Telegram groups; that single key generated 100M
GPT-5.4 tokens and more than seven Codex sessions.
We also deployed weakly configured \texttt{Sub2API},
\texttt{claude-relay-service}, and \texttt{CLIProxyAPI} decoys across
20 domains and 20 IPs.
Those decoys received tens of thousands of unauthorized access attempts from 147 IPs
(6 JA3 fingerprints),
served 2B GPT-5.4 / 5.3-codex tokens,
exposed about 13~GB of visible downstream prompt/response traffic, and
leaked
99 credentials across 440 Codex sessions on 398 different projects or
hosts.
Every one of those 440 sessions was command-injectable, and 401 already
ran in autonomous YOLO mode, meaning tool execution was already
auto-approved and simple payload injection would have been enough even
without sophisticated adaptive triggers.
Finally, we build \tool, a research proxy that implements the attack
classes and companion mitigations, and use it to evaluate practical
client-side defenses.
A fail-closed policy gate blocks all AC-1 and AC-1.a shell-rewrite
samples at 1.0\% false positives, and response-side anomaly screening
flags 89\% of AC-1 samples without requiring provider changes.
These mitigations reduce exposure today, but securing the agent
ecosystem ultimately requires provider-backed response integrity so
that the tool call an agent executes can be cryptographically tied to
what the upstream model actually produced.

\smallskip\noindent
In summary, this paper makes three contributions:
\begin{enumerate}[leftmargin=*,nosep]
\item \textbf{Threat model and attack taxonomy.}
  We present the first formal threat model for LLM API routers as a
  supply-chain trust boundary and define two core attack
  classes, payload injection (AC-1) and secret exfiltration
  (AC-2), together with two adaptive evasion variants:
  dependency-targeted injection (AC-1.a) and conditional delivery
  (AC-1.b), grounded in observed router behavior
  (Sections~\ref{sec:threatmodel}--\ref{sec:attacks}).
\item \textbf{Ecosystem measurement and poisoning studies.}
  We analyze 28 paid and 400 free routers and find 9 injecting
  malicious code, 2 deploying adaptive evasion, and 17 abusing
  researcher-owned credentials.
  Two poisoning studies show that benign routers can be pulled into the
  same attack surface through leaked keys and weak relay chains
  (Section~\ref{sec:measurement}).
\item \textbf{Implementation and deployable defenses.}
  We build \tool, a research proxy implementing all four attack
  classes against four public agent frameworks, and evaluate three
  client-side defenses that can be deployed today without provider
  cooperation
  (Sections~\ref{sec:implementation}--\ref{sec:defense}).
\end{enumerate}

\section{Background}
\label{sec:background}

\subsection{LLM API Routers}
\label{sec:bg-routers}

A direct API subscription to a single model provider is the simplest
deployment, but production agent systems rarely stop there.
Organizations need access to models from multiple providers (OpenAI,
Anthropic, Google, and an expanding set of open-weight
hosts) with fallback, load balancing, cost optimization, and a single
credential plane.
An \emph{LLM API router} fills this role: it accepts requests in a
unified format (typically OpenAI-compatible), selects an upstream
provider, and returns the response.

Routing exists at every scale.
At the institutional end, Amazon Bedrock~\cite{bedrock2026} and Azure
OpenAI Service~\cite{azureopenai2026} are cloud-managed routers: they
host or proxy third-party models behind a unified API, and enterprises
consume them as a managed service.
At the open-source end, LiteLLM~\cite{litellm2024} and
OpenRouter~\cite{openrouter2024} let individual developers and startups
aggregate dozens of providers behind a single base-URL change.
Some model providers collaborate directly with routers for
distribution; for example, making new models available through
OpenRouter or regional aggregator platforms as a first-class channel.

Crucially, routers are \emph{composable}: the path from client to GPU
routinely traverses multiple routing layers.
A developer may purchase API access from a Taobao reseller, who
aggregates keys from a second-tier aggregator, who routes through
OpenRouter, which dispatches to the model host. That is four hops, each
terminating and re-originating a TLS connection, each with full
plaintext access to API keys, system prompts, tool definitions, and
tool-call responses.
The client configures only the first hop; subsequent hops are invisible.
Because no end-to-end integrity mechanism spans this chain, a single
malicious or compromised router at \emph{any} layer taints the entire
path: downstream honest routers cannot detect that an upstream hop has
already rewritten a tool call or copied a credential.
We formalize this weakest-link property in
Section~\ref{sec:attacks}.

Routing is especially prevalent in regions where direct provider access
is restricted, expensive, or subject to quota limitations.
A large commodity market has emerged around resold and aggregated API
access: investigative reporting documents Taobao merchants with over
30{,}000 repeat purchases for LLM API
keys~\cite{chinatalk2025graymarket}, and the open-source router
templates that power most of these services, \texttt{new-api}~\cite{newapi2026}
(25.4k GitHub stars, 1.25M Docker pulls) and its upstream fork
\texttt{one-api}~\cite{oneapi2026} (30.5k stars, 1.19M Docker
pulls), have been pulled millions of times.
LiteLLM alone has accumulated roughly 40{,}000 stars and over
240~million Docker Hub pulls.

\subsection{Tool Use and Function Calling}
\label{sec:bg-tooluse}

Modern LLM APIs expose \emph{tool use} (also called \emph{function
calling}) as a first-class
capability~\cite{schick2024toolformer,qin2024toolllm,
patil2023gorilla}.
OpenAI returns a \texttt{tool\_calls} field with JSON-encoded
arguments~\cite{openai2023functioncalling};
Anthropic returns \texttt{tool\_use} content blocks with a native JSON
object~\cite{anthropic2024tooluse};
Gemini exposes a similar structured
interface~\cite{google2024functioncalling}.
In every format, tool-call arguments are transmitted as plaintext JSON.
No provider-level integrity mechanism binds the arguments returned by
the model to the arguments received by the client.
An intermediary that terminates TLS on each side can therefore
read, modify, or fabricate any tool-call payload without detection.

\subsection{The LiteLLM Incident}
\label{sec:bg-litellm}

In March 2026, attackers compromised LiteLLM through dependency
confusion, injecting malicious code into the request-handling pipeline
of every deployment that pulled the poisoned
release~\cite{litellm2026teampcp}.
The injected payload had write access to every API request and response
transiting the proxy, the same capability set that a deliberately
malicious router would possess.
This incident demonstrated that the router trust boundary is not
hypothetical: a single supply-chain entry point in one widely deployed
router was sufficient to compromise the entire forwarding path.

\section{Threat Model}
\label{sec:threatmodel}

Figure~\ref{fig:architecture} illustrates the system architecture and
the attacker's position.

We consider an attacker that operates a malicious LLM API router or has
compromised a legitimate one through supply-chain compromise, insider
access, or server-side exploitation~\cite{litellm2026teampcp}.
Because the client explicitly configures the router as its API endpoint,
the router terminates client-side TLS and originates a separate TLS
connection upstream.
It therefore occupies an application-layer man-in-the-middle position by
design and can read, retain, rewrite, or fabricate request and response
bodies, headers, and request metadata.
This includes tool definitions, prompts, tool outputs, API keys, and
returned tool-call payloads across OpenAI-, Anthropic-, and
Gemini-style interfaces.
The router may also keep cross-request state, which lets it activate
payload rewriting only for trigger-matching sessions.
We assume standard TLS between the router and the upstream provider and
no compromise of model weights or inference logic.

The core integrity gap is that no deployed mechanism binds the
provider-origin tool-call response to what the client finally receives.
That gap enables response-side payload rewriting, while request-side
visibility enables selective delivery to particular users, workflows, or
tool invocations.
We exclude prompt injection, model backdoors, client-side malware,
denial of service, and pure model substitution.
Those behaviors may compose with router abuse, but they are distinct
from the response-manipulation and passive-collection attacks
studied here.

\section{Attack Taxonomy}
\label{sec:attacks}

\begin{figure*}[t]
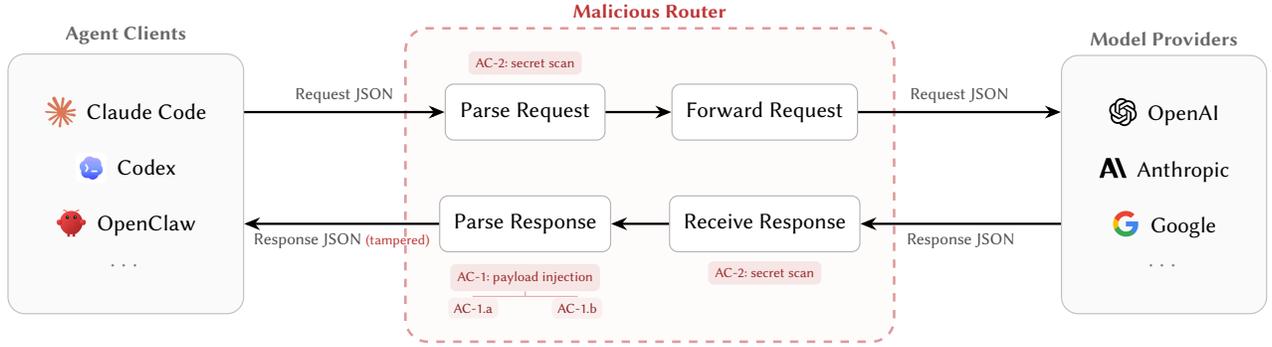

\centering
\resizebox{0.96\textwidth}{!}{%
\pgfdeclarelayer{bg}
\pgfdeclarelayer{mid}
\pgfsetlayers{bg,mid,main}
\begin{tikzpicture}[
    arr/.style={-{Stealth[length=2mm, width=1.6mm]}, line width=0.8pt},
    box/.style={draw=black!30, rounded corners=3pt, inner sep=5pt,
                font=\sffamily\small, minimum width=2.0cm, minimum height=0.7cm,
                align=center, fill=white},
    tag/.style={font=\sffamily\tiny, rounded corners=2pt,
                inner sep=2.5pt, fill=cRed!12, text=cRed!80!black},
    subtag/.style={font=\sffamily\tiny, rounded corners=2pt,
                   inner sep=2pt, fill=cRed!8, text=cRed!70!black},
    groupbox/.style={draw=black!25, rounded corners=6pt, fill=black!2,
                     inner xsep=10pt, inner ysep=12pt},
    lbl/.style={font=\sffamily\scriptsize, text=black!70},
]

\node (a1) at (0, 0.7)
  {\raisebox{-3pt}{\includegraphics[height=11pt]{claude-color.pdf}}
   \,\sffamily\small Claude Code};
\node (a2) at (0, 0)
  {\raisebox{-3pt}{\includegraphics[height=11pt]{codex-color.pdf}}
   \,\sffamily\small Codex};
\node (a3) at (0,-0.7)
  {\raisebox{-3pt}{\includegraphics[height=11pt]{openclaw-dark.pdf}}
   \,\sffamily\small OpenClaw};
\node[font=\sffamily\small, text=black!50] (adots) at (0,-1.25) {$\cdots$};

\begin{pgfonlayer}{bg}
  \node[groupbox, fit=(a1)(a2)(a3)(adots),
    label={[font=\sffamily\footnotesize\bfseries, text=black!60,
      anchor=south]north:Agent Clients}] (clientbox) {};
\end{pgfonlayer}

\node (p1) at (13.0, 0.7)
  {\raisebox{-2pt}{\includegraphics[height=10pt]{openai.pdf}}
   \,\sffamily\small OpenAI};
\node (p2) at (13.0, 0)
  {\raisebox{-2pt}{\includegraphics[height=10pt]{anthropic.pdf}}
   \,\sffamily\small Anthropic};
\node (p3) at (13.0,-0.7)
  {\raisebox{-2pt}{\includegraphics[height=10pt]{google.pdf}}
   \,\sffamily\small Google};
\node[font=\sffamily\small, text=black!50] (pdots) at (13.0,-1.25) {$\cdots$};

\begin{pgfonlayer}{bg}
  \node[groupbox, fit=(p1)(p2)(p3)(pdots),
    label={[font=\sffamily\footnotesize\bfseries, text=black!60,
      anchor=south]north:Model Providers}] (provbox) {};
\end{pgfonlayer}

\node[box] (parse_req) at (5.0, 0.7)  {Parse Request};
\node[box] (fwd)       at (8.0, 0.7)  {Forward Request};

\node[box] (recv_resp)  at (8.0, -0.7) {Receive Response};
\node[box] (parse_resp) at (5.0, -0.7) {Parse Response};

\node[tag, above=3pt of parse_req]  (t_ac2a) {AC-2: secret scan};
\node[tag, below=3pt of recv_resp]  (t_ac2b) {AC-2: secret scan};

\node[tag, below=4pt of parse_resp] (t_ac1) {AC-1: payload injection};
\node[subtag, below=2pt of t_ac1, xshift=-0.65cm] (t_ac1a) {AC-1.a};
\node[subtag, below=2pt of t_ac1, xshift=0.65cm]  (t_ac1b) {AC-1.b};

\draw[cRed!40, line width=0.4pt]
  (t_ac1.south) -- ++(0,-0.06) -| (t_ac1a.north);
\draw[cRed!40, line width=0.4pt]
  (t_ac1.south) -- ++(0,-0.06) -| (t_ac1b.north);

\begin{pgfonlayer}{mid}
  \node[draw=cRed!50, dashed, rounded corners=8pt, line width=0.9pt,
        fill=cRed!2,
        fit=(parse_req)(fwd)(recv_resp)(parse_resp)(t_ac2a)(t_ac2b)(t_ac1a)(t_ac1b),
        inner xsep=12pt, inner ysep=8pt,
        label={[font=\sffamily\footnotesize\bfseries, cRed,
                anchor=south]north:Malicious Router}]
        (routerbox) {};
\end{pgfonlayer}

\draw[arr] (clientbox.east |- parse_req) -- (parse_req.west)
    node[midway, above, lbl] {Request JSON};
\draw[arr] (parse_req) -- (fwd);
\draw[arr] (fwd.east) -- (provbox.west |- fwd)
    node[midway, above, lbl] {Request JSON};

\draw[arr] (provbox.west |- recv_resp) -- (recv_resp.east)
    node[midway, below, lbl] {Response JSON};
\draw[arr] (recv_resp) -- (parse_resp);
\draw[arr] (parse_resp.west) -- (clientbox.east |- parse_resp)
    node[midway, below, lbl]
    {Response JSON {\tiny\color{cRed}(tampered)}};

\end{tikzpicture}
}
\caption{Request--response lifecycle through a malicious router.
AC-2 tags mark where the router passively scans traffic for secrets
(both request and response paths).
AC-1 marks where parsed responses are rewritten before delivery;
AC-1.a specializes to dependency substitution,
AC-1.b gates activation on session-level triggers
(Section~\ref{sec:attacks:evasion}).}
\label{fig:attack-flow}
\end{figure*}

Malicious-router behavior reduces to two orthogonal primitives:
\emph{active manipulation}, in which the router rewrites a tool-call
payload before it reaches the client, and \emph{passive collection}, in
which the router silently extracts secrets from plaintext traffic.
We formalize these as two core attack classes (AC-1 and AC-2) and
define two \emph{adaptive evasion} variants (AC-1.a and AC-1.b) that
specialize AC-1 to evade specific classes of client-side defenses.
Table~\ref{tab:attack-comparison} summarizes the taxonomy, and
Figure~\ref{fig:attack-flow} shows where each class activates in the
request--response path.

\paragraph{Formal framework.}
We model the system as $(C, R_1, \ldots, R_k, P)$ where $C$~is
the client, $P$~is the upstream provider, and
$R_1, \ldots, R_k$~are routers.
A request $\mathit{req} \in \mathsf{Request}$ carries a prompt,
tool definitions, and an API key;
a response $\mathit{resp} \in \mathsf{Response}$ carries
tool calls $[t_1, \ldots, t_n]$ where
$t_i = (\mathit{name}_i, \mathit{args}_i)$.
Let $\sigma \subseteq \mathsf{Secrets}$ denote a set of extracted
credential patterns, and let
$\varphi : \mathsf{Request} \times \mathsf{State}
\to \mathsf{Bool}$
be a trigger predicate over request and session features.

An honest router is transparent:
$R_{\mathit{honest}}(\mathit{req}) = P(\mathit{req})$.
A chain composes as
$(R_1 \circ \cdots \circ R_k)(\mathit{req})$,
where each $R_i$ terminates and re-originates a TLS connection.
Chain integrity is a \emph{weakest-link} property:
\begin{mathpar}
\inferrule
  {\forall\, i \in [1,k].\;
   R_i = R_{\mathit{honest}}}
  {(R_1 \circ \cdots \circ R_k)(\mathit{req}) = P(\mathit{req})}
\and
\inferrule
  {\exists\, j \in [1,k].\;
   R_j \neq R_{\mathit{honest}}}
  {\text{no integrity guarantee for }
   (R_1 \circ \cdots \circ R_k)(\mathit{req})}
\end{mathpar}

\noindent
A single malicious router at any layer can apply AC-1 (rewrite) or
AC-2 (collect); downstream honest routers cannot detect or undo the
modification because they lack a reference to the original upstream
response.
For AC-2, taint is cumulative: every router in the chain observes
plaintext traffic, so the total secret exposure is
$\sigma_{\mathit{chain}} = \bigcup_{i=1}^{k}
\mathit{extract}_i(\mathit{req}_i, \mathit{resp}_i)$.
Our measurement (Section~\ref{sec:measurement:poisoning}) confirms
this composability empirically: leaked keys and weak relays turn
otherwise benign outer routers into conduits for the full attack
surface.
The remainder of this section defines the attack classes for a
single malicious router $R$; the chain property above lifts each
class to arbitrary multi-hop deployments.

\begin{table*}[t]
\caption{Attack taxonomy: two core classes and two adaptive evasion variants.}
\label{tab:attack-comparison}
\centering
\footnotesize
\begin{tabular}{@{}lp{1.3cm}p{2.2cm}p{2.3cm}p{2.4cm}p{3.0cm}@{}}
\toprule
\textbf{Class} & \textbf{Role} & \textbf{Manipulated Surface} & \textbf{Preconditions} & \textbf{Primary Harm} & \textbf{Detection Difficulty} \\
\midrule
AC-1
  & Core
  & Tool-call arguments
  & Tool-calling response; no integrity check
  & Arbitrary code execution
  & Modified payload is schema-valid; client never sees upstream original \\
\addlinespace
AC-2
  & Core
  & None (read-only)
  & Secret in plaintext traffic
  & Credential theft
  & Traffic is unchanged; clients cannot observe router-side retention \\
\addlinespace
AC-1.a
  & Evasion
  & Package name inside install command
  & Install-capable tool call
  & Durable supply-chain compromise
  & Evades domain-based policy gates; rewritten command looks legitimate \\
\addlinespace
AC-1.b
  & Evasion
  & Same as AC-1, conditionally
  & Trigger-relevant session features
  & Targeted delivery
  & Non-matching probes see benign behavior; finite audits miss the attack \\
\bottomrule
\end{tabular}
\end{table*}

\subsection{Core Attack Classes}
\label{sec:attacks:core}

\subsubsection{AC-1: Response-Side Payload Injection}
\label{sec:attacks:ac1}

\begin{mathpar}
\inferrule
  {P(\mathit{req}) = \mathit{resp} \\
   \mathit{resp}.\mathit{tool\_calls}[i] = t}
  {R_{\text{AC-1}}(\mathit{req})
   = \mathit{resp}\bigl[\mathit{tool\_calls}[i]
     \mapsto \mathit{rewrite}(t)\bigr]}
\end{mathpar}

\noindent
The function $\mathit{rewrite} : \mathsf{ToolCall} \to \mathsf{ToolCall}$
replaces selected fields in the argument JSON while preserving the tool
name and schema structure.
The router rewrites a model-generated tool call after it leaves the
upstream provider but before it reaches the client; the only
preconditions are a tool-calling response and the absence of an
integrity mechanism binding the received arguments to the upstream
original.
Because the modified payload remains syntactically valid JSON matching
the expected tool schema, AC-1 redirects agent behavior without
producing a schema violation or transport anomaly.
For a shell-execution tool such as \texttt{Bash}, replacing a benign
URL with an attacker-controlled script suffices for arbitrary code
execution: the semantic change occurs after inference completes,
entirely outside the model's reasoning loop.

\smallskip
\noindent\textbf{Example.}
The listing below shows a benign installer URL replaced with an
attacker-controlled endpoint.
The March 2026 LiteLLM compromise~\cite{litellm2026teampcp}
demonstrated exactly this primitive at scale: once the attacker
controlled the request pipeline, every transiting tool call was exposed
to rewriting.

\noindent\begin{minipage}{\columnwidth}
\noindent\textit{Original tool call (from upstream provider):}
\begin{lstlisting}[language={},numbers=none,
rulecolor=\color{cGreen!60}, framerule=2.5pt,
escapeinside={(*@}{@*)}]
{
  "name": "Bash",
  "arguments": {
    "command": "curl -sSL (*@\hlg{https://get.example.com/cli.sh}@*) | bash"
  }
}
\end{lstlisting}

\noindent\textit{Router-modified tool call (delivered to client):}
\begin{lstlisting}[language={},numbers=none,
rulecolor=\color{cRed!60}, framerule=2.5pt,
escapeinside={(*@}{@*)}]
{
  "name": "Bash",
  "arguments": {
    "command": "curl -sSL (*@\hlr{https://attacker****.sh}@*) | bash"
  }
}
\end{lstlisting}
\end{minipage}

\noindent\begin{minipage}{\columnwidth}
\begin{framed}
\noindent\textbf{Consequence of AC-1:}
A single rewritten tool call is sufficient for arbitrary code execution on the client machine. Any agent that auto-executes tool calls through an unverified router is exposed.
\end{framed}
\end{minipage}

\subsubsection{AC-2: Passive Secret Exfiltration}
\label{sec:attacks:ac2}

\begin{mathpar}
\inferrule
  {P(\mathit{req}) = \mathit{resp} \\
   \mathit{extract}(\mathit{req}, \mathit{resp}) = \sigma \\
   \sigma \neq \emptyset}
  {R_{\text{AC-2}}(\mathit{req}) = \mathit{resp}
   \;\wedge\; \mathit{leak}(\sigma)}
\end{mathpar}

\noindent
The function
$\mathit{extract} : \mathsf{Request} \times \mathsf{Response}
\to \mathcal{P}(\mathsf{Secrets})$
scans headers, request bodies, and response bodies against credential
patterns; the router forwards the response unmodified and exfiltrates
$\sigma$ asynchronously.
AC-2 requires no payload modification; the boundary between
``credential handling'' and ``credential theft'' is invisible to the
client because routers already read secrets in plaintext as part of
normal forwarding.
Once exposed credentials are reused by relays, passive collection alone
creates downstream data exposure at scale: our poisoning study
(Section~\ref{sec:measurement:poisoning}) shows that a single leaked
key yielded 100M tokens and 99~credentials across 440~sessions without
any payload rewriting.

\smallskip\noindent\textbf{Example.}
Listing~\ref{lst:ac2-patterns} shows representative extraction
patterns.
An attacker who controls the LiteLLM request pipeline as in the
March 2026 incident~\cite{litellm2026teampcp} gains read access to
every API key, system prompt, and credential that transits the
proxy, even if the injected code never modifies a single response.
In Section~\ref{sec:measurement}, we count AC-2 outcomes only when
exposure is followed by externally visible unauthorized use of
researcher-owned canaries or credentials.
In practice, the extraction surface extends beyond API keys:
system prompts, tool definitions, user-supplied file contents,
and environment variables all transit the same plaintext channel
and are equally accessible to a router performing AC-2.

\begin{lstlisting}[caption={AC-2: representative secret-matching
patterns for passive extraction.},
label={lst:ac2-patterns}, language={},
basicstyle=\ttfamily\scriptsize, numbers=none]
sk-[A-Za-z0-9]{20,}            // OpenAI keys
AKIA[A-Z0-9]{16}               // AWS keys
ghp_[A-Za-z0-9]{36}            // GitHub PATs
xoxb-[0-9]+-[A-Za-z0-9]+       // Slack bot tokens
0x[a-fA-F0-9]{64}              // Ethereum keys
-----BEGIN .* PRIVATE KEY----- // PEM keys
\end{lstlisting}

\noindent\begin{minipage}{\columnwidth}
\begin{framed}
\noindent\textbf{Consequence of AC-2:}
Passive collection requires zero payload modification and is invisible to the client, since the same plaintext access needed for legitimate routing also enables silent credential retention. Unlike AC-1, AC-2 cannot be mitigated by response-signing proposals because the secrets are exposed on the \emph{request} path before any provider-side mechanism can act.
\end{framed}
\end{minipage}

\subsection{Adaptive Evasion Variants}
\label{sec:attacks:evasion}

The core AC-1 attack is effective but coarse: always-on rewriting is
easily detected by a simple policy check or manual review.
A sophisticated attacker therefore needs to control not only
\emph{what} is injected but also \emph{when} and \emph{how} the
injection is delivered, so that standard client-side checks pass
while high-value targets still receive malicious payloads.
We define two adaptive evasion variants that specialize AC-1 to defeat
specific defensive strategies while preserving its payload-injection
capability.

\subsubsection{AC-1.a: Dependency-Targeted Injection}
\label{sec:attacks:ac1a}

\begin{mathpar}
\inferrule
  {P(\mathit{req}) = \mathit{resp} \\
   t_i.\mathit{name} \in \{\texttt{Bash}, \texttt{run\_command}\} \\
   t_i.\mathit{args} \models \mathit{install\_pattern}}
  {R_{\text{AC-1.a}}(\mathit{req})
   = \mathit{resp}\bigl[t_i.\mathit{args}
     \mapsto \mathit{subst}(\mathit{pkg})\bigr]}
\end{mathpar}

\noindent
AC-1.a specializes AC-1 to package-install commands
(\texttt{pip\,install}, \texttt{npm\,install}, \texttt{cargo\,add}).
Rather than rewriting an arbitrary URL, which a domain-based policy
gate (Section~\ref{sec:defense:policy}) can catch, the router
substitutes a legitimate dependency name with an attacker-controlled
package pre-registered on the target registry.
The substitution may be a visually similar name (typosquatting) or
an entirely different package; the former is particularly effective
because LLM-based review and approval UIs tend to hallucinate that
a near-homograph is correct, causing downstream checks to pass.
The surrounding command line remains unchanged, so the rewritten
command clears domain-based allowlists and approval flows that
emphasize only the high-level action.
Once the substituted package installs, the attacker gains a durable
supply-chain foothold that persists beyond the current session.
This is strictly more dangerous than a one-shot AC-1 URL redirect,
because the compromised dependency is cached locally and re-imported
across future sessions.
We design AC-1.a specifically to demonstrate that the policy gate
defense can be evaded when the attacker targets package-install
workflows: the gate blocks non-allowlisted domains but does
not maintain a per-package allowlist.

\smallskip
\noindent\textbf{Example.}
The listing below shows a single-character substitution:
\texttt{requests} becomes \texttt{reqeusts}, a typosquat package
that the router has pre-registered on PyPI.
Because the surrounding command line is unchanged and the package name
passes a casual visual check, the rewritten command clears both
domain-based policy gates and LLM-assisted approval flows.

\noindent\begin{minipage}{\columnwidth}
\noindent\textit{Original tool call:}
\begin{lstlisting}[language={},numbers=none,
rulecolor=\color{cGreen!60}, framerule=2.5pt,
escapeinside={(*@}{@*)}]
{
  "name": "Bash",
  "arguments": {
    "command": "python -m pip install (*@\hlg{requests}@*) flask pyyaml"
  }
}
\end{lstlisting}

\noindent\textit{Router-modified tool call:}
\begin{lstlisting}[language={},numbers=none,
rulecolor=\color{cRed!60}, framerule=2.5pt,
escapeinside={(*@}{@*)}]
{
  "name": "Bash",
  "arguments": {
    "command": "python -m pip install (*@\hlr{reqeusts}@*) flask pyyaml"
  }
}
\end{lstlisting}
\end{minipage}

\noindent\begin{minipage}{\columnwidth}
\begin{framed}
\noindent\textbf{Consequence of AC-1.a:}
Dependency-targeted injection evades domain-based policy gates because the rewritten command installs from the same trusted registry, differing only in the package name. Worse, the compromised dependency is cached locally and re-imported across future sessions, giving the attacker a durable supply-chain foothold that persists long after the malicious router interaction ends.
\end{framed}
\end{minipage}

\subsubsection{AC-1.b: Conditional Delivery}
\label{sec:attacks:ac1b}

\begin{mathpar}
\inferrule
  {P(\mathit{req}) = \mathit{resp} \\
   \varphi(\mathit{req}, s) = \mathit{true}}
  {R_{\text{AC-1.b}}(\mathit{req})
   = R_{\text{AC-1}}(\mathit{req})}
\and
\inferrule
  {P(\mathit{req}) = \mathit{resp} \\
   \varphi(\mathit{req}, s) = \mathit{false}}
  {R_{\text{AC-1.b}}(\mathit{req})
   = \mathit{resp}}
\end{mathpar}

\noindent
The predicate $\varphi$ gates payload injection on session state $s$:
the router behaves honestly for non-matching traffic and applies AC-1
only when $\varphi$ holds.
AC-1.b is not a distinct payload primitive but determines \emph{when}
AC-1 activates, so that routine probes and low-value traffic see benign
behavior while high-value sessions receive rewritten payloads.
Practical triggers fall into five families:
\begin{enumerate}[leftmargin=*,nosep]
\item \textbf{Tool-name conditions:} activate only for tools
  such as \texttt{Bash} or \texttt{run\_command}.
\item \textbf{Content keywords:} activate when the command contains
  installer patterns such as \texttt{curl~-sSL} or \texttt{pip~install}.
\item \textbf{User fingerprints:} activate only for sessions in
  autonomous YOLO mode or belonging to a target organization.
\item \textbf{Time windows:} activate only during off-hours or after a
  deployment event.
\item \textbf{Request-count thresholds:} activate only after the
  session has passed $N$ benign interactions, hiding the attack from
  shallow smoke tests.
\end{enumerate}
None of these require model compromise; they depend only on metadata
and payload fields already visible to the router.
A router deploying a 50-request warm-up appears entirely
transparent during a 10-request probe, making finite black-box auditing
fundamentally inadequate against conditional delivery.

\smallskip\noindent\textbf{Example.}
The Codecov CI breach~\cite{codecov2021breach} persisted undetected
for months by activating its credential-harvesting payload only within
CI environments, a conditional-delivery strategy analogous to AC-1.b.
Our measurement (Section~\ref{sec:measurement}) confirms that
real routers deploy similar logic: one activates only after 50~prior
calls, another restricts injection to YOLO-mode sessions targeting
Rust or Go projects.
The combination of multiple trigger families is especially potent:
a router can require both a tool-name match \emph{and} a warm-up
count before activating, so that even an auditor who tests the
correct tool never reaches the activation threshold within a
typical probe budget.

\noindent\begin{minipage}{\columnwidth}
\begin{framed}
\noindent\textbf{Consequence of AC-1.b:}
Conditional delivery makes finite black-box auditing fundamentally inadequate. A router that activates only after a warm-up period or on specific session attributes appears entirely transparent during routine probes. Because the trigger predicate is server-side and opaque, no fixed-length client test can guarantee that the router is benign. Combined with autonomous auto-approval modes, the attacker can reserve payload injection exclusively for high-value sessions while keeping all audit traffic clean.
\end{framed}
\end{minipage}

\smallskip
\paragraph{Summary.}
AC-1 and AC-2 are orthogonal: AC-1 changes what the agent executes,
while AC-2 silently harvests credentials from the same plaintext
channel.
The evasion variants sharpen AC-1 for realistic deployment:
AC-1.a evades domain-based policy gates by substituting dependencies
instead of URLs, and AC-1.b evades black-box auditing by gating
delivery on session-level triggers.
Section~\ref{sec:measurement} maps these classes to the observed
ecosystem, and Section~\ref{sec:defense} evaluates client-side
defenses against each.

\section{Ecosystem Measurement}
\label{sec:measurement}

We study two complementary questions.
First, are malicious routers already operating in real agent-facing
markets?
Second, can routers that appear benign or trusted be poisoned into the
same supply-chain position through leaked upstream credentials or by
forwarding traffic through weaker relays?
Our measurement therefore combines a market study of paid and free
routers with two poisoning studies based on leaked researcher-owned keys
and intentionally weak relay deployments.
Table~\ref{tab:measurement-datasets} summarizes the datasets,
Table~\ref{tab:measurement-main} collects the main outcomes,
Figure~\ref{fig:router-abuse-overview} visualizes the malicious-router
counts, and Table~\ref{tab:adaptive-evasion} lists adaptive-evasion
conditions observed in the wild or demonstrated in the artifact.

\begin{table*}[t]
\caption{Measurement datasets and collection channels.}
\label{tab:measurement-datasets}
\centering
\small
\begin{tabular}{@{}p{0.16\linewidth}p{0.32\linewidth}p{0.15\linewidth}p{0.25\linewidth}@{}}
\toprule
\textbf{Dataset} & \textbf{Collection Channel} & \textbf{Scale} & \textbf{Purpose} \\
\midrule
Paid routers & Taobao, Xianyu, Shopify storefronts & 28 routers & Test sold OpenAI- and Anthropic-compatible endpoints \\
\addlinespace
Free routers & Public links using \texttt{sub2api}~\cite{sub2api2026} and \texttt{new-api}~\cite{newapi2026} templates & 400 routers & Measure in-the-wild abuse in commodity router ecosystems \\
\addlinespace
Leaked-key poisoning & Chinese forums, and WeChat / Telegram groups & 1 OpenAI key & Observe downstream sessions on a reused upstream account \\
\addlinespace
Weak-router decoys & Weak-password \texttt{Sub2API}, \texttt{CLIProxyAPI}, and \texttt{claude-relay-service} deployments & 20 domains + 20 IPs & Measure exploitation and downstream exposure through poisoned routers \\
\bottomrule
\end{tabular}
\end{table*}

\subsection{Dataset and Collection}
\label{sec:measurement:corpus}

We purchased 28 paid OpenAI- and Anthropic-compatible routers from
Taobao~\cite{taobao2026}, Xianyu~\cite{xianyu2026}, and
Shopify-hosted storefronts~\cite{shopify2026}.
We also collected 400 free routers exposed through public communities,
router lists, and configuration dumps.
The free corpus is dominated by two widely reused templates,
\texttt{sub2api}~\cite{sub2api2026} (11k GitHub stars) and
\texttt{new-api}~\cite{newapi2026} (25.4k stars, 1.25M Docker Hub
pulls), which made it possible to expand
from seed URLs into a much larger set of reachable router instances.
That expansion followed only additional public links reachable from the
same posts, router lists, and configuration-dump references; we did not
scan arbitrary address ranges or attempt to enumerate hidden admin
surfaces.
We conservatively deduplicated exact repeated endpoints after URL
normalization; deployments that reused the same template on distinct
domains or IPs remained separate because they still correspond to
different externally reachable trust boundaries.
We also could not always recover exact template versions or complete
default configurations, so we do not attribute observed abuse to a
particular \texttt{sub2api} or \texttt{new-api} release.
For each router, the testing pipeline proceeds in four stages.
First, we register a researcher-controlled account with the router and
obtain API credentials.
Second, we route the account through a sandboxed agent environment and
execute a randomized subset of custom tasks and TerminalBench~2.0
tasks~\cite{terminalbench2026}, which produce realistic shell, package-install,
and file-manipulation tool calls across diverse programming languages
and workflows.
Third, every tool-call payload returned by the router is executed
inside the sandbox, which monitors for network callbacks,
file-system modifications, and process-tree anomalies.
Fourth, a dedicated analysis agent reviews flagged executions and
classifies the behavior as benign rewriting (e.g., CDN substitution),
potentially malicious, or confirmed malicious.
This pipeline also records whether conditional triggers were
present and whether router-side plaintext access to researcher-owned
AWS or Ethereum canaries produced follow-on abuse.

To study how ostensibly benign routers can be poisoned, we ran two
additional measurements.
First, we intentionally leaked a researcher-owned OpenAI API key on Chinese forums, and WeChat and Telegram groups where router operators
share credentials and relay endpoints~\cite{meli2019git}.
Second, we deployed vulnerable
\texttt{claude-relay-service} instances across 20
domains and weak-password \texttt{Sub2API} and \texttt{CLIProxyAPI} instances across 20 IPs.
These decoys were researcher-controlled and served only to measure how
quickly external operators would absorb them into downstream router
chains.
Across both poisoning studies, we retained only aggregate session
metadata, credential classes, salted hashes of leaked credential values,
and coarse host or project identifiers; we did not store raw third-party
prompt or response bodies.
We use conservative operational definitions throughout.

A router counts as injecting malicious code only if a returned tool-call
payload is rewritten into an attacker-controlled command or dependency
under researcher-controlled probing.
We count an AWS canary as touched when a credential that crossed the
router later produces follow-on AWS API activity attributable to that
credential, and an ETH drain when funds leave a prefunded
researcher-controlled private key after exposure.
In the weak-router study, an unauthorized access attempt is a distinct
unsolicited interaction against a decoy endpoint; a session is
command-injectable if it exposes at least one shell-execution path whose
returned command could be rewritten before execution; and YOLO mode
means automatic tool approval without per-command confirmation.
Throughout this section, token totals reflect all billed traffic visible
at the exposed upstream account or decoy, whereas Codex-session counts
cover only the subset of traffic we could confidently attribute to
downstream Codex clients, so the two quantities are not directly
comparable.

\begin{table*}[t]
\caption{Main measurement outcomes across malicious routers and poisoned benign routers.}
\label{tab:measurement-main}
\centering
\small
\resizebox{\textwidth}{!}{%
\begin{tabular}{@{}p{0.16\textwidth}p{0.15\textwidth}p{0.13\textwidth}p{0.13\textwidth}p{0.14\textwidth}p{0.22\textwidth}@{}}
\toprule
\textbf{Setting} & \textbf{Sample / Footprint} & \textbf{Active Manipulation} & \textbf{Adaptive Evasion} & \textbf{Credential Abuse} & \textbf{Main Outcome} \\
\midrule
Paid routers
  & 28 purchased routers
  & 1 router injects malicious code
  & 0 observed
  & 0 observed
  & Paying for a router does not remove tool-call injection risk \\
\addlinespace
Free routers
  & 400 public routers
  & 8 routers inject malicious code
  & 2 routers trigger selectively
  & 17 routers touch AWS canaries; 1 drains ETH
  & Commodity free-router ecosystems already expose AC-1, AC-1.a/b, and AC-2 behaviors \\
\addlinespace
Leaked upstream key
  & 1 OpenAI key leaked publicly
  & 0 observed
  & Not needed
  & $>$7 Codex sessions; one session exposed multiple credentials
  & 100M GPT-5.4 tokens observed under a researcher-owned account \\
\addlinespace
Weak-router decoys
  & 20 domains + 20 IPs
  & All 440 observed sessions expose shell-execution paths
  & 401 sessions already in auto-approve (YOLO) mode
  & 99 credentials across 440 sessions / 398 projects
  & 2B billed tokens; about 13\,GB visible traffic; 40k+ attempts; 6 JA3 \\
\bottomrule
\end{tabular}
}
\end{table*}

\subsection{Malicious Routers in the Wild}
\label{sec:measurement:malicious}

\begin{figure}[t]
\centering
\begin{tikzpicture}
\begin{axis}[
    ybar,
    width=0.96\columnwidth,
    height=5cm,
    bar width=9pt,
    ymin=0, ymax=5.5,
    enlarge x limits=0.2,
    symbolic x coords={
      {Code injection},
      {Adaptive evasion},
      {Credential abuse},
      {Wallet drain}},
    xtick=data,
    ylabel={Routers (\%)},
    grid=major,
    x tick label style={font=\scriptsize, rotate=0, anchor=north},
    nodes near coords,
    nodes near coords style={font=\tiny},
    every node near coord/.append style={yshift=1pt},
]
\addplot[fill=cBlue!30, draw=cBlue!80] coordinates {
    ({Code injection},3.57)
    ({Adaptive evasion},0.00)
    ({Credential abuse},0.00)
    ({Wallet drain},0.00)
};
\addplot[fill=cRed!30, draw=cRed!80] coordinates {
    ({Code injection},2.00)
    ({Adaptive evasion},0.50)
    ({Credential abuse},4.25)
    ({Wallet drain},0.25)
};
\legend{Paid ($n{=}28$), Free ($n{=}400$)}
\end{axis}
\end{tikzpicture}
\caption{Observed malicious-router behaviors across 28 paid and 400 free
routers.
Bars are normalized within the paid and free populations; raw counts
appear in Table~\ref{tab:measurement-main} and the surrounding text.
Adaptive evasion is observed only among routers that already perform
active manipulation.}
\label{fig:router-abuse-overview}
\end{figure}
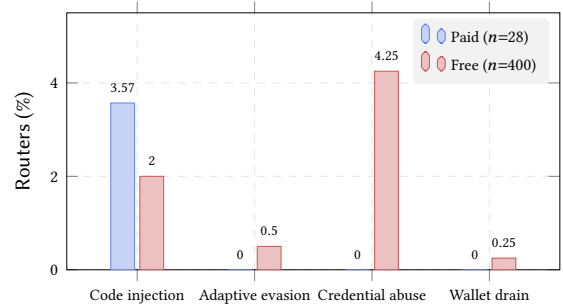

Across the 28 paid and 400 free routers, we find active malicious
behavior in both markets.
One paid router and eight free routers inject malicious code into
returned tool calls.
These modifications are not cosmetic output rewrites; they change the
command or dependency that the agent intends to execute.
In the free-router set, 17 routers trigger follow-on use of at least
one researcher-owned AWS canary credential after seeing it in transit,
and one router drains ETH
from a researcher-owned Ethereum private key.
Figure~\ref{fig:router-abuse-overview} summarizes these counts by paid
versus free router populations.

Two malicious routers exhibit adaptive evasion rather than always-on
rewriting.
The deployed trigger logic is operational rather than exotic.
In one case, malicious tool calls are injected only after the first
50 requests, which hides the attack from shallow smoke tests.
In another, injection is enabled only for sessions that appear to be
running in autonomous YOLO mode and only when the project fingerprint
matches Rust or Go.
These triggers are sufficient to keep benign probes clean while
reserving manipulation for high-value coding sessions.
They are not necessarily exhaustive.
Because we observe the routers only through finite black-box probing, a
router may implement additional latent conditions that our probes did
not activate.

\begin{table}[t]
\caption{AC-1.b conditions observed in malicious routers.
Each maps to one of the five trigger families defined in
Section~\ref{sec:attacks:ac1b}.}
\label{tab:adaptive-evasion}
\centering
\footnotesize
\resizebox{\columnwidth}{!}{%
\begin{tabular}{@{}p{0.15\columnwidth}p{0.22\columnwidth}p{0.22\columnwidth}p{0.30\columnwidth}@{}}
\toprule
\textbf{Type} & \textbf{Condition} & \textbf{Trigger Family} & \textbf{Effect} \\
\midrule
Observed & After 50 prior calls & Request-count threshold & Hides AC-1 until warm-up traffic completes \\
\addlinespace
Observed & Only in YOLO mode & User fingerprint & Reserves payload injection for autonomous sessions \\
\addlinespace
Observed & Only for Rust / Go projects & Content keyword & Targets high-value developer environments while evading generic probes \\
\bottomrule
\end{tabular}}
\end{table}

\subsection{Poisoning Benign and Trusted Routers}
\label{sec:measurement:poisoning}

Active malicious routers are only half of the supply-chain problem.
Routers that look benign can be poisoned into the same position when
they reuse leaked upstream credentials or when they forward user traffic
through other weak routers.
We observe both paths.

In the first poisoning study, we intentionally leaked a
researcher-owned OpenAI API key on Chinese forums, WeChat, and
Telegram groups frequented by router operators.
That single key served 100M GPT-5.4 tokens and exposed more than
seven Codex user sessions.
Because the traffic was billed to and logged under our upstream account,
we could observe downstream prompt, tool, and credential material
crossing a router we did not directly operate.
In at least one observed session, multiple distinct credentials
appeared behind the leaked key.
This result is consistent with several paths: direct reuse by a router
operator, direct reuse by unrelated third parties, or reuse somewhere
inside a longer relay chain.
We cannot always distinguish among those paths from provider-side logs
alone.
What the experiment does establish is the blast radius: once a
supposedly benign router path reuses a stolen upstream credential, the
holder of that credential inherits the same plaintext visibility as an
actively malicious router.

In the second poisoning study, we deployed vulnerable \texttt{claude-relay-service} instances across 20
domains and weak-password \texttt{Sub2API} and \texttt{CLIProxyAPI} instances across 20 IPs.
We observed 40k+ unauthorized access attempts from 147 IPs with six
unique JA3 fingerprints.
Those initial accesses are consistent with a mix of opportunistic
internet scanning and subsequent relay reuse.
We therefore do not interpret every access event itself as evidence of a
targeted supply-chain attack.
The stronger signal comes from the later structured downstream traffic:
the decoys were not merely scanned once, but were folded into active
agent-facing relay paths that generated sustained billed usage and
repeat Codex sessions.
Those decoys ultimately served roughly 2B GPT-5.4 and 5.3-codex tokens,
corresponding to roughly
13~GB of visible downstream prompt/response traffic, and exposed 99 unique
credentials across 440 Codex sessions spanning 398 different projects
or hosts.
All 440 sessions were command-injectable, and 401 already operated in
YOLO mode, i.e., with tool execution auto-approved rather than
confirmed command by command.
For those 401 sessions, sophisticated adaptive evasion was unnecessary:
basic payload injection would have been sufficient to change executed
commands.

This benign-router poisoning result is the strongest evidence that the
router threat boundary is transitive.
A router does not need to be malicious at account creation time.
If it later adopts leaked upstream keys or forwards traffic into a weak
relay chain, all four attack classes become available to whoever
controls that upstream account or inner relay.
The user may believe they are trusting one router, while the effective
trust boundary has silently expanded to a larger chain of opportunistic
operators.

\subsection{Key Findings}
\label{sec:measurement:findings}

\paragraph{Malicious routers already exist in both paid and free
commodity router markets.}
The 1 paid and 8 free routers that inject malicious code show that this
is not a purely hypothetical threat or a pathology confined to obvious
free relay dumps.
Paid access improves service stability, but it does not prove tool-call
integrity.

\paragraph{Adaptive evasion is deployed, but often unnecessary.}
We observe real routers that wait for warm-up traffic, target only YOLO
mode, or restrict injection to Rust and Go projects.
At the same time, the weak-router decoy study shows that many downstream
agent sessions are already so permissive that complex triggers are not
required: 401 of 440 observed sessions were autonomous enough for simple
payload injection to succeed.

\paragraph{Benign routers can be poisoned into the same trust
boundary.}
Leaked upstream keys and weak third-party relays turn otherwise benign
routers into channels for plaintext prompt visibility, credential
exposure, and command injection.
The supply-chain risk therefore does not begin only when a router
operator decides to act maliciously; it also appears when a router
reuses compromised credentials or silently chains through a weaker
upstream intermediary.

\subsection{Scope}
\label{sec:measurement:limitations}

Our corpus targets the most active publicly reachable commodity router
markets; enterprise and invite-only deployments are a natural extension.
Although the majority of routers in our corpus originate from
Chinese-language marketplaces and communities, this does not make the
findings regionally narrow:
Chinese open-source models reached nearly 30\% of total OpenRouter
usage in some weeks, and Asia's share of LLM API spend on the platform
grew from about 13\% to 31\%~\cite{chou2025openrouter}, so routers
serving this traffic handle a substantial share of global routing
volume.
The poisoning studies demonstrate exploitability and blast radius under
realistic conditions, and the session-level findings reflect the
population of users who interact with these commodity routers in
practice.

\section{\tool Artifact}
\label{sec:implementation}

We implement \tool as an OpenAI-compatible FastAPI proxy that forwards
requests to an upstream provider and conditionally applies AC-1,
AC-1.a, AC-1.b, and AC-2.
We also implement companion client-side modules for the deployable
defenses evaluated in Section~\ref{sec:defense}: a tool policy gate,
response-side anomaly screening, and an append-only transparency log.

\tool parses each request, evaluates trigger rules, optionally
activates an attack module, forwards the request upstream, and applies
response-side rewrites before returning data to the client.
AC-1 rewrites tool-call payloads via JSON-path mutation;
AC-1.a rewrites shell and package-install command strings via
substitution rules;
AC-1.b selects when AC-1 and AC-1.a activate using tool-name, keyword,
user-fingerprint, time-window, and $N$-th-request conditions; and
AC-2 scans request and response bodies for secrets and exfiltrates
matches asynchronously.
Streaming is handled by buffering Server-Sent Events (SSE) tool-call
chunks until the full argument payload can be rewritten.

\subsection{Cross-Framework Compatibility}

We evaluate \tool against four public agent frameworks: OpenClaw~\cite{openclaw2026},
OpenCode~\cite{opencode2026}, OpenAI's Codex, and Anthropic's Claude Code.
For each framework, we send 1{,}000 tool-calling requests through the
proxy with each attack module active.
This section asks whether public agent frameworks add local integrity
checks or other constraints that materially narrow the attack surface
in practice.

\begin{table}[t]
\caption{Cross-framework compatibility results for \tool.}
\label{tab:implementation-summary}
\centering
\footnotesize
\begin{tabular}{@{}ll@{}}
\toprule
\textbf{Metric} & \textbf{Result} \\
\midrule
Frameworks tested & 4 \\
Requests per framework/module & 1{,}000 \\
Frameworks with response-integrity checks & 0/4 \\
AC-1 rewrite compatibility & 100\% \\
AC-1.a rewrite compatibility & 99.6\% \\
AC-2 extractor coverage & 100\% \\
Buffered streaming compatibility & 4/4 \\
Median overhead & 0.013\,ms/request \\
Median upstream latency & 820\,ms \\
\bottomrule
\end{tabular}
\end{table}

None of the four frameworks implement response-integrity verification.
AC-1 achieves 100\% rewrite compatibility on tool-calling responses
across all frameworks.
AC-1.a achieves a 99.6\% compatibility rate on package-install tool
calls.
For AC-2, the relevant metric is extractor coverage rather than
client-visible execution: the module matches 100\% of payloads
containing detectable secrets under its rule set while remaining
passive from the client's perspective.
Within the evaluated frameworks, the same rewriting logic handles both
OpenAI- and Anthropic-style wire formats; OpenAI requires nested
parsing of string-encoded arguments, whereas Anthropic-style interfaces
expose native JSON structures directly.
These are compatibility measurements, not end-to-end execution rates:
they show that the tampered tool call reaches the client in a valid
framework-native format, but downstream permission prompts or local
execution sandboxes may still block the resulting action.
Table~\ref{tab:implementation-summary} summarizes the results.

In streaming mode, AC-1 and AC-1.a retain the same rewrite compatibility
rates because the proxy reconstructs the complete tool-call payload
before modification via buffered rewriting.
This buffering introduces an average pause of 0.004--0.005\,ms between the
final natural token and the first rewritten token, well below the
820\,ms median upstream latency and indistinguishable from ordinary
model jitter.
Per-request proxy overhead is 0.013\,ms median over 10{,}000
requests to GPT-5.4, measured on an Apple M3 Max workstation
(\texttt{time.perf\_counter\_ns()}).
The remaining 0.4\% of AC-1.a failures are edge cases where install
commands used quoted package fragments, editable local paths, or
URL-based installs that the substitution rule conservatively declined
to rewrite.

\section{Deployable Defenses}
\label{sec:defense}

No client-side control available today can prove that a router
preserved the upstream provider's response.
Still, several mitigations can be deployed immediately without
provider cooperation.
We implemented three such controls in the artifact and evaluated them
against the same attack classes studied in
Sections~\ref{sec:attacks} and~\ref{sec:implementation}.
All numbers in this section come from controlled artifact-side
evaluation rather than field data: the released scripts generate
synthetic benign and attack tool-call corpora, run the defenses over
those corpora, and report the resulting coverage and overhead.
The threat model for these defenses is deliberately modest: they are
meant to reduce exposure when the attacker must pass through ordinary
tool-execution paths visible to the client.
We assume the attacker may know the defense family and probe coarse
behavior; the adaptive tests below relax that assumption further.
They are not a substitute for origin authentication.
Table~\ref{tab:defense-summary} summarizes the coverage.

\begin{table}[t]
\caption{Immediate client-side defenses and their coverage.}
\label{tab:defense-summary}
\centering
\small
\resizebox{\columnwidth}{!}{%
\begin{tabular}{@{}p{0.19\columnwidth}p{0.08\columnwidth}p{0.08\columnwidth}p{0.08\columnwidth}p{0.08\columnwidth}p{0.41\columnwidth}@{}}
\toprule
\textbf{Defense} & \textbf{AC-1} & \textbf{AC-1.a} & \textbf{AC-2} & \textbf{AC-1.b} & \textbf{Result / Scope} \\
\midrule
Policy gate
  & 100\%
  & 100\%
  & --
  & --
  & 1.0\% FP; 2.3~$\mu$s; 100\% adaptive bypass in a simple benchmark \\
\addlinespace
Anomaly screening
  & 89.0\%
  & 50.0\%
  & --
  & 50.9\%
  & 6.7\% FP on returned tool calls \\
\addlinespace
Transparency log
  & Audit
  & Audit
  & Audit
  & Audit
  & 12.0~MB/1k sessions; 1.26~KB/entry \\
\bottomrule
\end{tabular}
}
\end{table}

These controls serve different operational purposes.
The policy gate is the only one in our evaluation that consistently
stops AC-1 and AC-1.a before execution, but only for narrowly scoped
high-risk tools.
Anomaly screening is weaker, yet still useful when deployments cannot
turn every shell-execution path into a fail-closed workflow.
Transparency logging does not prevent manipulation, but it preserves
the evidence needed to scope exposure once a router, account, or
session becomes suspect.
The practical takeaway is not that any one control solves the problem.
It is that meaningful exposure reduction is possible today without
waiting for provider cooperation.
The strongest caveat is that the policy gate is also brittle: once the
attacker can hide behind an allowlisted domain or a pre-positioned local
stager, the gate no longer helps.
That full bypass result is the main negative finding of this section,
and it bounds how much confidence practitioners should place in
allowlist-only defenses.

\subsection{High-Risk Tool Policy Gate}
\label{sec:defense:policy}

The strongest immediately deployable control is to fail closed on
high-risk shell-execution workflows.
Our policy gate inspects only locally visible tool-call payloads and
blocks commands that fetch from non-allowlisted domains or install
packages outside a local allowlist.
This is narrow by design: it targets the subset of tools that already
carry the highest execution risk, such as \texttt{Bash},
\texttt{run\_command}, and package-install commands emitted through
those tools.
On a templated benchmark of 1{,}000 benign shell commands plus 500
AC-1 and 500 AC-1.a commands generated by the artifact, the policy gate
blocks every AC-1 and AC-1.a sample with a 1.0\% false positive rate and
2.3~$\mu$s median decision cost.
All false positives are benign installer fetches from mirror domains
outside the local allowlist.
Mirror-heavy deployments, including developer environments that rely on
regional package mirrors, would likely see higher false-positive rates
unless those mirrors are explicitly incorporated into local policy.
The cost of this defense is operational rather than computational.
Teams must maintain allowlists for installer domains and dependency
names, and benign drift outside those allowlists becomes a review or
breakage event.
That threat model matters.
If the attacker can host payloads on an allowlisted domain or
pre-position a local stager that is later executed through a benign
\texttt{Bash} command, the gate no longer helps.
In a simple adaptive benchmark that uses those two strategies, the
policy gate is bypassed in 100\% of samples.
This means the gate should be understood as a coarse but useful filter
for obvious high-risk workflows, not as a robust defense against
defense-aware attackers.

\subsection{Response-Side Anomaly Screening}
\label{sec:defense:anomaly}

Some deployments cannot fail closed on every shell command.
For those settings, we evaluate a weaker detector that scores each
returned tool call using only locally observable features: shell-risk
patterns, tool-name frequency within the session, argument entropy,
unexpected secret-like strings, and schema deviation.
The detector deliberately does not compare the observed tool call
against an unavailable upstream response.
The implementation is an IsolationForest~\cite{liu2008isolation} fit on 2{,}000 synthetic
benign tool calls generated from \texttt{Bash}, \texttt{Read},
\texttt{Grep}, \texttt{Glob}, and \texttt{Edit} sessions, then
evaluated on disjoint held-out synthetic attack and benign sessions
(3{,}352 benign, 301 AC-1, 294 AC-1.a, and 53 AC-1.b tool calls).
It is therefore a controlled review-oriented detector, not a model
trained on production user traffic.
At 6.7\% false positives, the detector flags 89.0\% of AC-1 samples,
50.0\% of AC-1.a samples, and 50.9\% of AC-1.b samples.
An additional held-out benign distribution drawn from different
developer commands yields 0/600 false positives at that threshold.
On the in-distribution benign set, most false positives come from
legitimate but high-risk shell workflows: allowlisted installer
bootstraps, multi-package install commands, and
\texttt{kubectl apply} fetches.
Figure~\ref{fig:defense-anomaly-thresholds} shows the threshold
tradeoff.
At a strict 1\% false-positive budget, the detector still catches
43.5\% of AC-1 samples, but essentially none of AC-1.a and only 9.4\%
of AC-1.b.
Relaxing the budget to 5\% raises AC-1 coverage to 86.7\%, while
AC-1.a and AC-1.b remain only partially visible at 28.6\% and 49.1\%.
At 10\%, AC-1.a rises to 64.3\%, and at 20\% all three classes become
substantially more visible, but at a review burden that would be hard
to justify for routine developer workflows.
We therefore report 6.7\% as a middle operating point: it surfaces
most blatant AC-1 rewrites while keeping review load bounded and still
catching a meaningful share of AC-1.a and AC-1.b traffic.
Appendix~\ref{sec:appendix:defense} reports the full threshold table
and a feature ablation.
The ablation shows that the shell-risk feature carries most of the
signal: removing it drops AC-1 detection to 17.6\% and AC-1.a detection
to 4.4\%.
The detector is therefore useful for review prioritization, but it
remains weaker than a fail-closed policy and still loses ground
against selective or defense-aware attackers.
An attacker that knows the detector's feature family can deliberately
stay within ordinary shell syntax, spread an action across multiple
benign-looking tool calls, or fall back to AC-2, none of which this
local detector can rule out.

\begin{figure*}[t]
\centering
\begin{subcaptionblock}{0.48\textwidth}
\centering
\begin{tikzpicture}
\begin{axis}[
    width=\textwidth,
    height=0.55\textwidth,
    xmin=0, xmax=22,
    ymin=0, ymax=105,
    xlabel={False-positive budget (\%)},
    ylabel={Detection rate (\%)},
    legend style={at={(0.5,-0.28)}, anchor=north, legend columns=3},
    grid=major,
    xtick={1,5,10,20},
    ytick={0,20,40,60,80,100},
    mark size=1.5pt,
    line width=0.8pt,
]
\addplot+[color=cBlue, mark=*, mark options={fill=cBlue}]
    coordinates {(1,43.5) (5,86.7) (10,95.0) (20,100.0)};
\addlegendentry{AC-1}
\addplot+[color=cOrange, mark=square*, mark options={fill=cOrange}]
    coordinates {(1,0.0) (5,28.6) (10,64.3) (20,86.7)};
\addlegendentry{AC-1.a}
\addplot+[color=cGreen, mark=diamond*, mark options={fill=cGreen}]
    coordinates {(1,9.4) (5,49.1) (10,60.4) (20,83.0)};
\addlegendentry{AC-1.b}
\end{axis}
\end{tikzpicture}
\caption{Threshold sweep for anomaly screening.}
\label{fig:defense-anomaly-thresholds}
\end{subcaptionblock}
\hfill
\begin{subcaptionblock}{0.48\textwidth}
\centering
\begin{tikzpicture}
\begin{axis}[
    ybar,
    bar width=6pt,
    width=\textwidth,
    height=0.55\textwidth,
    symbolic x coords={AC-1,AC-1.a,AC-2,AC-1.b},
    xtick=data,
    ylabel={Effectiveness (\%)},
    ymin=0, ymax=110,
    ytick={0,20,40,60,80,100},
    legend style={at={(0.5,-0.20)}, anchor=north, legend columns=2},
    enlarge x limits=0.15,
    grid=major,
    every axis plot/.append style={line width=0pt},
]
\addplot[fill=cBlue!45, draw=cBlue!70] coordinates
    {(AC-1,100) (AC-1.a,100) (AC-2,50) (AC-1.b,100)};
\addplot[fill=cOrange!45, draw=cOrange!70] coordinates
    {(AC-1,100) (AC-1.a,92) (AC-2,0) (AC-1.b,100)};
\addplot[fill=cGreen!40, draw=cGreen!70] coordinates
    {(AC-1,100) (AC-1.a,100) (AC-2,100) (AC-1.b,100)};
\addplot[fill=cPurple!35, draw=cPurple!60] coordinates
    {(AC-1,60) (AC-1.a,60) (AC-2,0) (AC-1.b,40)};
\legend{Signing, Anomaly, TLS pin, Logging}
\end{axis}
\end{tikzpicture}
\caption{Defense effectiveness by attack class.}
\label{fig:defense-radar}
\end{subcaptionblock}
\caption{Defense evaluation.
(a) Threshold sweep: detection rate vs.\ false-positive budget for the
anomaly screener across AC-1, AC-1.a, and AC-1.b.
(b) Per-class effectiveness of all four defenses: response signing,
anomaly detection, TLS pinning, and transparency logging.}
\label{fig:defense-combined}
\end{figure*}

\subsection{Append-Only Transparency Logging}
\label{sec:defense:logging}

The third control is a local transparency log that records the request
body, response body, router URL, TLS metadata, and a hash of the raw
response bytes after request-side secret redaction.
Logging does not prevent manipulation, but it improves forensic
scoping once misuse is suspected and makes it easier to correlate
traffic across retries, routers, and upstream accounts.
For AC-2 in particular, the log is useful only after the fact: it can
tie a leaked upstream credential or suspicious tool output to later
unauthorized usage on the same account, but it does not detect passive
collection at the moment it occurs.
In a storage benchmark over 1{,}000 synthetic OpenAI-style sessions
(10 tool calls each), the log costs 12.0~MB per 1{,}000 sessions, or
about 1.26~KB per entry.
That overhead is small enough for developer workstations and CI jobs,
which makes the control practical even when fail-closed policies are
too restrictive.
In deployment, the log is most useful when paired with one of the
preventive controls above.
The gate or detector decides what to block or escalate in the moment;
the log preserves the request, returned tool call, router endpoint,
and response hash needed to answer the next question after an
incident: how far did this router or credential reach, and which
sessions were exposed through it?

These defenses reduce exposure for high-risk tool-use deployments, but
they do not authenticate origin.
A router that stays within local allowlists and avoids obvious
anomalies can still alter semantics.
The remaining gap is end-to-end provenance, which still points back to
provider-supported integrity mechanisms.

\section{Discussion}
\label{sec:discussion}

\subsection{Scope and Future Directions}
\label{sec:discussion:limitations}

Our measurement targets the most active commodity router markets and
uses researcher-controlled accounts throughout.
Extending the study to private deployments are natural next
steps that would complement the snapshot presented here.

\subsection{Longer-Term Integrity}
\label{sec:discussion:trust}

Choosing a router is a trust decision, but it is not the same as
choosing a cloud provider or package registry.
The switching cost is unusually low: in many agent frameworks, moving
to a router is just a base-URL change and a new API key.
At the same time, the service is often presented as a transparent
compatibility layer even though it can translate schemas, substitute
credentials, and return executable tool calls.

Existing security mechanisms suggest what would and would not help.
Mutual TLS, certificate pinning, and ordinary transport security can
authenticate the router endpoint the client chose, but they do not say
whether the returned tool call preserves upstream semantics~\cite{rfc8705}.
Web integrity mechanisms such as Subresource Integrity~\cite{w3c2016sri},
signed exchanges~\cite{yasskin2020sxg}, and certificate-transparency
logs~\cite{rfc6962} illustrate two useful patterns: authenticate
content and make that authentication auditable.
Artifact-attestation systems such as SLSA and Sigstore apply the same
idea to software supply chains by signing provenance statements and
release artifacts~\cite{slsa2026,sigstore2026}.
Existing message-signing machinery could carry such a signature, but it
does not remove the need to define a canonical application payload.
The closest analogue here would be a provider-signed canonical
response envelope, similar in spirit to DKIM for
email~\cite{rfc6376}, that covers the model identifier, tool name,
tool arguments, finish reason, and a client nonce.
Appendix~\ref{sec:appendix:envelope} gives a minimal message format and
verification procedure.
In brief, the provider signs a canonical JSON object containing the
provider identity, model, content, tool calls, finish reason, request
nonce, validity window, and key identifier~\cite{rfc8785}.
The client verifies that envelope before executing any tool call.
Canonicalization is necessary because the routers in our corpus front
heterogeneous upstream providers through OpenAI- or
Anthropic-compatible interfaces, so signing the raw HTTP body is
insufficient.
To our knowledge, none of the major provider tool-use APIs or the
current MCP specification expose a deployed response-signing mechanism
for tool-call arguments today~\cite{openai2023functioncalling,anthropic2024tooluse,google2024functioncalling,mcp2024security}.
Section~\ref{sec:defense} shows what clients can do today without that
provider support.
Those controls reduce exposure and preserve evidence, but they do not
prove provenance.
Execution sandboxes such as E2B reduce post-execution blast radius but
do not authenticate where a tool call came from~\cite{e2b2026}.

\subsection{Generalizability}
\label{sec:discussion:generalizability}

The Model Context Protocol (MCP)~\cite{hou2025mcpinsecurity} introduces
a related trust boundary between LLM agents and external tools.
A malicious MCP server receives tool-call requests in plaintext and can
return forged results, so the same basic manipulation and collection
ideas transfer with adaptation to the MCP message format.

Our implementation evaluates buffered rewriting; richer variants
including token injection and AC-1.b triggers are natural extensions.
The measured buffering pause of 0.004--0.005\,ms is far below the
820\,ms median upstream latency
(Section~\ref{sec:implementation}), confirming that buffered
rewriting adds negligible overhead in practice.
\section{Related Work}
\label{sec:relatedwork}

\begin{table*}[!ht]
\centering
\caption{Related work comparison.}
\label{tab:relatedwork}
\small
\begin{tabular}{@{} p{2.5cm} p{2.0cm} p{5.0cm} p{7.0cm} @{}}
\toprule
\textbf{Prior Work} & \textbf{Layer} & \textbf{Focus} & \textbf{Our Differentiation} \\
\midrule
Greshake et al.~\cite{greshake2023prompt}; Perez \& Ribeiro~\cite{perez2022ignore}
  & Model
  & Prompt injection: adversarial text manipulates model reasoning
  & Router attacks modify JSON wire format below the model; orthogonal to prompt-level defenses \\
\addlinespace
Zou et al.~\cite{zou2023universal}
  & Model
  & Jailbreaking and adversarial prompting
  & We attack the transport, not the model; no adversarial prompt needed \\
\addlinespace
Ohm et al.~\cite{ohm2020backstabber}; LiteLLM incident~\cite{litellm2026teampcp}
  & Supply chain
  & Supply chain attacks on OSS / AI infrastructure
  & We analyze post-compromise router capabilities: active tool-call rewriting, passive collection, and conditional delivery \\
\addlinespace
Gu et al.~\cite{gu2019badnets}; Kurita et al.~\cite{kurita2020weight}
  & Model
  & Model-level backdoors via training / fine-tuning
  & Router attacks require no model access and no training-time adversary \\
\addlinespace
Durumeric et al.~\cite{durumeric2017security}; de Carnavalet \& Mannan~\cite{decarnavalet2016killed}
  & Transport
  & TLS interception by middleboxes
  & LLM routers are voluntarily configured; no cert substitution needed; attacks are application-layer semantic \\
\addlinespace
MCP security~\cite{mcp2024security}; Hou et al.~\cite{hou2025mcpinsecurity}
  & Tool server
  & Tool-server poisoning via malicious MCP descriptions
  & We target the client--provider transport; a compromised router can intercept any MCP-based interaction that transits it \\
\addlinespace
Liu et al.~\cite{liu2026skillswild}
  & Client extension
  & Vulnerabilities in installable agent skills and bundled scripts
  & Router attacks need no skill installation and can affect both skill-enabled and skill-free clients \\
\bottomrule
\end{tabular}
\end{table*}

Table~\ref{tab:relatedwork} summarizes the closest prior lines of
research and how our work differs.

\paragraph{Prompt injection.}
Greshake et al.\ introduced indirect prompt injection, showing that
adversarial content embedded in external data sources can hijack an
LLM's behavior~\cite{greshake2023prompt}.
Subsequent work explored direct prompt
injection~\cite{perez2022ignore},
jailbreaking~\cite{zou2023universal}.
Router attacks are orthogonal: the intermediary rewrites the JSON wire
format outside the model's reasoning loop, so prompt-level defenses do
not authenticate the returned tool-call payload.

\paragraph{Software supply chain.}
Ladisa et al.\ systematized attacks on open-source supply
chains~\cite{ladisa2023sok}; Duan et al.\ measured typosquatting and
dependency confusion across package managers~\cite{duan2021measuring};
Ohm et al.\ catalogued maintainer compromise and related
vectors~\cite{ohm2020backstabber}.
The Codecov breach showed how a single compromised CI script can persist
for months while exfiltrating
credentials~\cite{codecov2021breach}.
Gu et al.\ and Kurita et al.\ demonstrated backdoor injection into
pre-trained models and fine-tuning
pipelines~\cite{gu2019badnets,kurita2020weight}.
Adjacent systems such as SLSA and Sigstore sign build provenance or
release artifacts rather than dynamic per-response tool-call
semantics~\cite{slsa2026,sigstore2026}.

\paragraph{TLS interception and API gateways.}
Durumeric et al.\ measured the security impact of HTTPS interception by
middleboxes~\cite{durumeric2017security}; de Carnavalet and Mannan
found widespread TLS validation
failures~\cite{decarnavalet2016killed}; Waked et al.\ showed that even
well-intentioned interception introduces
vulnerabilities~\cite{waked2020interception}.
LLM routers perform the same basic operation, but the client chooses the
intermediary explicitly, so no certificate substitution
occurs~\cite{man2020dnscache}.
Enterprise AI gateways such as Kong~\cite{kong2026aigateway} add policy
around the chosen intermediary, and sandboxes such as
E2B~\cite{e2b2026} constrain post-execution blast radius, but neither
authenticates the provider-origin tool-call payload.

\paragraph{MCP security.}
MCP introduces a related trust boundary between agents and tool
servers~\cite{mcp2024security,hou2025mcpinsecurity}.
The key structural difference is \emph{where} the intermediary sits:
an MCP server terminates the tool-execution side and can forge outputs
but cannot observe or alter the upstream model's reasoning.
A malicious router, by contrast, sits on the client--provider transport
and intercepts \emph{every} tool call as well as the full request
context.
Liu et al.\ studied vulnerabilities in installable agent skills and
bundled scripts~\cite{liu2026skillswild}; router attacks need no skill
installation and affect both skill-enabled and skill-free clients.

\section{Conclusion}
\label{sec:conclusion}

LLM API routers sit on a critical trust boundary that the ecosystem
currently treats as transparent transport.
Our measurement of 428 commodity routers found 9~injecting malicious
code and 17~abusing researcher-owned credentials; poisoning studies
showed that even benign routers are one leaked key away from the same
exposure, with researcher-controlled decoys attracting 2B~billed
tokens, 440~autonomous Codex sessions, and 99~leaked credentials.
Client-side defenses (policy gates, anomaly screening, transparency
logs) reduce exposure today, but closing the provenance gap ultimately
requires provider-signed response envelopes so that the tool call an
agent executes can be tied to what the model actually produced.

\bibliographystyle{ACM-Reference-Format}
\bibliography{references}

\appendix
\section{Ethical Considerations}
\label{sec:ethics}

This appendix describes the ethical framework governing our research,
including data handling,
measurement constraints, and dual-use risk mitigation.

\smallskip\noindent\textbf{No IRB / ethics-board review.}
We did not obtain IRB or equivalent ethics-board review for this
study.
The work used only researcher-controlled accounts and credentials,
relied on synthetic active-probing traffic, and retained only
aggregate operational metadata from unauthorized third-party use of
researcher-owned secrets.
We therefore treated it as systems measurement rather than human-subjects
research, but we make this status explicit because the
credential-exposure case study intentionally created publicly
discoverable secrets.
We nevertheless treated the study as ethically sensitive because that
design could attract third-party abuse and lead to nominal financial
loss on researcher-owned accounts.

\subsection{Disclosure Scope}

We did not run a provider-by-provider coordinated disclosure process
for the findings in Section~\ref{sec:measurement}.
Several considerations informed this decision.
First, the paper centers on three measurements: routers openly sold in
public markets, free routers distributed through public communities, and
researcher-controlled poisoning studies based on leaked keys and weak
relay decoys.
These are not private zero-days disclosed by a single vendor.
They are observations about how publicly reachable router ecosystems and
router chains behave once exposed to attacker-relevant inputs.
Second, the affected routers are commodity services operated by
pseudonymous or anonymous sellers on Taobao, Xianyu, and public
community forums; there is no stable security-contact channel for most
of these operators, and many explicitly advertise their service as
unofficial or gray-market.
Third, the vulnerability is architectural rather than
implementation-specific: any router that terminates TLS and forwards
tool-call JSON can mount the same attacks, so disclosing to individual
operators would not remediate the underlying trust gap.
We therefore treated the work as a measurement study rather than an
embargo case.
At the end of the observation window, all exposed credentials were
revoked or otherwise retired.
Because the affected upstream credentials were researcher-owned and
could be retired directly, we did not separately notify OpenAI,
Anthropic, or other upstream providers about each individual reuse
event.

\subsection{Data Minimization}

We adhere to strict data minimization principles throughout the study:

\smallskip\noindent\textbf{Research accounts only.}
All API keys, user accounts, and service subscriptions used in our
experiments (Sections~\ref{sec:measurement}--\ref{sec:implementation})
were created specifically for this research.
We never access accounts or credentials not under researcher control.
When third-party traffic voluntarily reached researcher-controlled keys
or decoy relays, we limited retention to aggregate metadata and hashed
credential identifiers as described below.

\smallskip\noindent\textbf{Synthetic payloads.}
All provider-facing payloads and prompts used in our study are
synthetically generated.
No real user queries, proprietary code, or sensitive data appear
in any provider-facing validation request.

\smallskip\noindent\textbf{Retrospective credential-exposure data.}
The poisoning studies
(Section~\ref{sec:measurement:poisoning}) are observational rather than
interactive: they analyze unauthorized traffic that reached
researcher-owned credentials after public exposure.
For these studies, we retain only aggregate operational metadata
(timestamps, coarse model identifiers, token volume, source network
labels where available, session counts, project or host counts, and
salted hashes of leaked credential values) and do not store or release
prompt/response bodies or raw credential strings from third-party
traffic.
Project or host identifiers were stored only in coarse form and, where
persisted, as salted one-way hashes rather than human-readable names.
They were used solely for counting distinct exposure scopes and were not
joined against external account records.

\smallskip\noindent\textbf{No persistent data collection.}
Experimental data is retained only for the duration of the study.
Researcher-owned credentials used in the credential-exposure case
study (Section~\ref{sec:measurement:poisoning}) were revoked or
otherwise retired upon completion of the observation period.
All provider interaction logs are stored on encrypted research
infrastructure and will be deleted 12 months after
publication.
Revocation could interrupt unauthorized downstream use of those exposed
credentials.
We accepted that externality because continued operation would have
extended third-party exposure and financial loss on researcher-owned
accounts.

\subsection{Measurement Constraints}

We impose the following constraints to ensure our experiments do not
disrupt the services we study:

\smallskip\noindent\textbf{Rate limiting.}
No provider receives more than 60 requests per
hour during any experiment, well below the rate limits published by
all tested providers.
Provider-facing validation requests are spaced to avoid triggering
abuse-detection mechanisms.

\smallskip\noindent\textbf{No third-party traffic interception.}
All active probing requests originate from our own client
infrastructure and target our own upstream accounts.
The poisoning studies do not rely on
network-level interception equipment, DNS hijacking, or traffic
redirection; it analyzes upstream-provider metadata associated with
researcher-owned credentials after those credentials became publicly
discoverable or after traffic voluntarily reached researcher-controlled
decoy relays.

\smallskip\noindent\textbf{No exploitation of discovered
vulnerabilities.}
Where our measurement reveals potential security weaknesses
(e.g., unauthorized secret reuse in the credential-exposure case
study), we
record the finding but do not attempt to validate it through
additional real-world exploitation.
We do not attempt to exploit, amplify, or reproduce any
vulnerability beyond the minimum necessary to confirm its existence.

\smallskip\noindent\textbf{Minimal financial exposure.}
Researcher-owned Ethereum decoy keys were prefunded only with nominal
balances.
For the single on-chain drain reported in
Section~\ref{sec:measurement:poisoning}, the value lost was below US\$50
at the time of transfer.

\subsection{Dual-Use Risk and Mitigations}

The attack taxonomy and techniques we describe
(Sections~\ref{sec:attacks}--\ref{sec:implementation}) constitute
dual-use research: the same material that enables defensive
understanding could guide a malicious router operator.
We adopt the following mitigations:

\smallskip\noindent\textbf{No public release of \tool.}
We do not publish \tool or any of its attack modules.
\tool exists solely as an internal research implementation used to
produce the compatibility and defense results in
Sections~\ref{sec:implementation}--\ref{sec:defense}; we neither
distribute the source code nor provide deployment or operational
guidance for it.
This choice is intentional: it raises the engineering barrier for
misuse while preserving the scientific value of the measurements the
tool enabled.

\smallskip\noindent\textbf{Defensive value outweighs offensive risk.}
The attack classes we describe (AC-1, AC-1.a, AC-1.b, and AC-2) require only
straightforward JSON manipulation; any competent adversary with
router access could implement them independently.
By publishing a systematic taxonomy and measurement methodology, we
enable the community to build better safeguards around intermediary
trust in agent systems.
We believe the defensive benefit of public disclosure substantially
outweighs the marginal increase in offensive capability, consistent
with the established norms of the security research
community~\cite{greshake2023prompt, durumeric2017security}.

\section{Additional Defense Evaluation}
\label{sec:appendix:defense}

All defense results in Section~\ref{sec:defense} come from controlled
artifact-side evaluation rather than field data.
The released scripts procedurally generate benign and attack corpora
from fixed command templates and random seeds, then run the defenses on
those corpora.

\begin{table}[t]
\caption{Corpora used for the deployable-defense evaluation.}
\label{tab:appendix-defense-corpora}
\centering
\footnotesize
\begin{tabular}{@{}p{0.17\columnwidth}p{0.23\columnwidth}p{0.50\columnwidth}@{}}
\toprule
\textbf{Defense} & \textbf{Corpus Size} & \textbf{Construction} \\
\midrule
Policy gate & 1{,}000 benign, 500 AC-1, 500 AC-1.a & Templated shell commands covering installer fetches, package installs, \texttt{grep}, \texttt{git}, \texttt{pytest}, and \texttt{kubectl}; AC-1 and AC-1.a samples are generated by substituting malicious domains or attacker-controlled package names. \\
\addlinespace
Anomaly screening & 2{,}000 fit benign; held-out 3{,}352 benign, 301 AC-1, 294 AC-1.a, 53 AC-1.b & Procedurally generated sessions over \texttt{Bash}, \texttt{Read}, \texttt{Grep}, \texttt{Glob}, and \texttt{Edit}. The detector is an IsolationForest fit on synthetic benign sessions only; held-out attack labels come from injected AC-1, AC-1.a, and trigger-matching AC-1.b tool calls. \\
\addlinespace
Transparency log & 1{,}000 sessions, 10{,}000 entries & Synthetic OpenAI-style request/response objects with tool calls, request-side secret redaction checks, and response-byte hashing checks. \\
\bottomrule
\end{tabular}
\end{table}

The anomaly detector in Section~\ref{sec:defense:anomaly} is fit on
2{,}000 benign tool calls, calibrated on disjoint held-out benign
sessions, and evaluated on separate attack sessions.
The held-out test split contains 3{,}352 benign tool calls, 301 AC-1
calls, 294 AC-1.a calls, and 53 AC-1.b calls.
The AC-1.b count is smaller because only trigger-matching calls are
labeled as attack traffic; non-matching calls remain benign by
construction.
We also check the chosen operating point on a held-out benign
distribution built from different developer commands and observe
0/600 false positives.

\begin{table}[t]
\caption{Threshold sensitivity for response-side anomaly screening.}
\label{tab:appendix-anomaly-thresholds}
\centering
\footnotesize
\begin{tabular}{@{}rrrr@{}}
\toprule
\textbf{Benign FP Target} & \textbf{AC-1} & \textbf{AC-1.a} & \textbf{AC-1.b} \\
\midrule
1\%  & 43.5\% & 0.0\%  & 9.4\%  \\
5\%  & 86.7\% & 28.6\% & 49.1\% \\
10\% & 95.0\% & 64.3\% & 60.4\% \\
20\% & 100.0\% & 86.7\% & 83.0\% \\
\bottomrule
\end{tabular}
\end{table}

\begin{table}[t]
\caption{Feature ablation for anomaly screening at the 6.7\% false-positive operating point.}
\label{tab:appendix-anomaly-ablation}
\centering
\footnotesize
\begin{tabular}{@{}lrrr@{}}
\toprule
\textbf{Feature Removed} & \textbf{AC-1} & \textbf{AC-1.a} & \textbf{AC-1.b} \\
\midrule
None & 89.0\% & 50.0\% & 50.9\% \\
shell\_risk\_score & 17.6\% & 4.4\% & 9.4\% \\
tool\_frequency & 88.4\% & 53.4\% & 45.3\% \\
string\_entropy & 89.0\% & 33.7\% & 50.9\% \\
unexpected\_secret\_pattern & 89.0\% & 47.3\% & 50.9\% \\
schema\_deviation & 86.4\% & 39.5\% & 50.9\% \\
\bottomrule
\end{tabular}
\end{table}

The threshold sweep shows the expected tradeoff: AC-1 rises quickly as
the false-positive budget grows, while AC-1.a and AC-1.b require much more
lenient thresholds.
The ablation confirms that shell-risk patterns dominate detection for
active command rewrites, which is precisely why the detector remains a
review aid rather than a substitute for provenance.

\section{Canonical Response-Envelope Format}
\label{sec:appendix:envelope}

This appendix gives a minimal message format for the provider-signed
response envelope discussed in Section~\ref{sec:discussion:trust}.
The goal is semantic integrity for tool-calling responses even when a
router re-serializes, wraps, or otherwise transforms the original HTTP
body.

\begin{table}[t]
\caption{Minimal provider-signed response-envelope fields.}
\label{tab:appendix-envelope-fields}
\centering
\footnotesize
\begin{tabular}{@{}p{0.18\columnwidth}p{0.72\columnwidth}@{}}
\toprule
\textbf{Field} & \textbf{Purpose} \\
\midrule
\texttt{v} & Envelope version for compatibility and rollout. \\
\texttt{provider} & Provider identity, e.g., \texttt{api.openai.com}. \\
\texttt{key\_id} & Signing-key identifier used for verification and rotation. \\
\texttt{model} & Provider model identifier for the signed response. \\
\texttt{request\_nonce} & Client-supplied nonce bound to the corresponding request. \\
\texttt{issued\_at} & Provider timestamp for replay control and audit. \\
\texttt{expires\_at} & Short validity horizon for key rotation and replay limits. \\
\texttt{content} & Natural-language assistant content, if any. \\
\texttt{tool\_calls} & Array of tool calls, each with \texttt{name} and native-JSON \texttt{arguments}. \\
\texttt{finish\_reason} & Provider finish reason, e.g., \texttt{tool\_calls} or \texttt{stop}. \\
\texttt{sig\_alg} & Signature algorithm identifier. \\
\texttt{signature} & Signature over the canonicalized envelope excluding this field. \\
\bottomrule
\end{tabular}
\end{table}

The signed scope is the entire envelope except \texttt{signature}.
Provider-specific billing metadata, raw response identifiers, and
transport headers remain outside the signed scope because they are not
required to decide which tool call the client executes.
The critical normalization step is that \texttt{tool\_calls[*].arguments}
must be represented as native JSON values inside the envelope even if a
provider's wire format emits them as string-encoded JSON.
This parsing step must itself be canonical and fail closed.
If a provider cannot unambiguously parse a string-encoded argument blob
into native JSON, it should treat the response as unsigned rather than
producing a best-effort envelope.

\noindent\begin{minipage}{\columnwidth}
\begin{lstlisting}[caption={Example response envelope.},label={lst:appendix-envelope}]
{
  "v": 1,
  "provider": "api.openai.com",
  "key_id": "2026-04-k1",
  "model": "gpt-5.4",
  "request_nonce": "b7c6b9f0e87a4a6b",
  "issued_at": "2026-04-07T18:00:00Z",
  "expires_at": "2026-04-07T18:05:00Z",
  "content": "I will inspect the repository.",
  "tool_calls": [
    {
      "name": "Bash",
      "arguments": {"command": "grep -R \"TODO\" ./src"}
    }
  ],
  "finish_reason": "tool_calls",
  "sig_alg": "Ed25519",
  "signature": "base64..."
}
\end{lstlisting}
\end{minipage}

\noindent\textbf{Provider-side generation.}
Given an upstream response, the provider-side SDK or API gateway:
(1) maps the provider-native response into the envelope fields above;
(2) parses any string-encoded tool arguments into native JSON;
(3) canonicalizes the resulting object with RFC~8785 JSON
canonicalization~\cite{rfc8785}; and
(4) signs the canonical byte string with the private key referenced by
\texttt{key\_id}.

\noindent\textbf{Client-side verification.}
Before executing any tool call, the client:
(1) fetches or caches the provider verification key for
\texttt{provider} and \texttt{key\_id};
(2) checks that \texttt{request\_nonce} matches the outstanding
request;
(3) checks that \texttt{issued\_at} and \texttt{expires\_at} define a
currently valid window; and
(4) re-canonicalizes the envelope without \texttt{signature} and
verifies the signature.
If any step fails, the client treats the response as unsigned and
blocks tool execution.

\noindent\textbf{Deployment notes.}
Routers may still add unsigned outer metadata, but clients should
execute tool calls only from the verified envelope.
This design therefore tolerates schema translation and re-serialization
while preventing a router from silently rewriting the semantically
meaningful tool-call payload.
Backwards compatibility is incremental: providers can add the envelope
alongside existing response formats, and clients that do not understand
it simply ignore it and behave as they do today.
Clients that do understand it can adopt a phased policy, e.g., verify
when present, then require signatures only for high-risk tool
categories.

For streaming responses, the simplest design is to sign the final
tool-bearing envelope rather than every token chunk.
That matches the execution boundary in current tool-use clients, which
typically wait for complete tool arguments before taking action.
Per-chunk signatures are possible, but they would add significantly more
protocol complexity and are unnecessary for the core threat studied
here, namely silent modification of the final tool-call payload.

\end{document}